\documentclass[a4paper,fleqn]{cas-sc}

\usepackage[round, sort]{natbib}

\usepackage{amssymb}
\usepackage[utf8]{inputenc}
\usepackage[export]{adjustbox}
\usepackage{amsfonts}
\usepackage{amsmath, amssymb}
\usepackage{bm}

\newcommand{\rev}[1]{{\color{black}#1}}

\newcommand{\pdiff}[2]{\frac{\partial #1}{\partial #2}}

\newcommand{\mean}[1]{\langle #1 \rangle}

\newcommand{\ttt}[1]{\texttt{#1}}

\begin{document}

\let\WriteBookmarks\relax
\def\floatpagepagefraction{1}
\def\textpagefraction{.001}
\shorttitle{Flow dynamics in the closure region of an internal ship air cavity}
\shortauthors{T. Mukha and R.E. Bensow}

\title [mode = title]{Flow dynamics in the closure region of an internal ship air cavity}

\author[1]{Timofey Mukha}[orcid=0000-0002-2195-8408]
\ead{timofey@chalmers.se}
\cormark[1]

\credit{Conceptualization of this study, Methodology, Investigation, Visualization, Data curation, Writing - Original draft preparation}

\address[1]{Chalmers University of Technology, Department of Mechanics and Maritime Sciences, Hörsalsvägen 7A, SE-412 96 Gothenburg, Sweden}

\author[1]{Rickard E. Bensow}[orcid=0000-0002-8208-0619]
\ead{rickard.bensow@chalmers.se}
\cormark[2]

\credit{Conceptualization of this study, Funding acquisition, Project administration, Resources, Supervision, Writing - review \& editing}

\cortext[cor1]{Corresponding author}
\cortext[cor2]{Principal corresponding author}

\begin{abstract}
This work is dedicated to providing a detailed account of the flow dynamics in the closure region of an internal ship air cavity. A geometrically simple multiwave test cavity is considered, and a simulation of the flow is conducted using large-eddy simulation coupled with an algebraic Volume of Fluid interface capturing method. Results reveal that the flow in the closure region is highly unsteady and turbulent. The main cause of this is established to be the pressure gradient occurring due to the stagnation of the flow on the beach wall of the cavity. The pressure gradient leads to a steep incline in the mean location of the air-water interface, which, in turn, leads to the flow separating from it and forming a recirculation zone, in which air and water are mixed. The separated flow becomes turbulent, which further facilitates the mixing and entrainment of air. Swarms of air bubbles leak periodically. Upstream of the closure region, the phase and length of the wave are found to be well-predicted using existing approximations based on linear flow theory. However, for the corresponding prediction of the amplitude of the wave the agreement is worse, with the estimates under-predicting the simulation results.
\end{abstract}


\maketitle

\section{Introduction}
Approximately 60\% of a ship's propulsion power is spent on overcoming friction drag~\citep{Makiharju2012}.
Consequently, development of innovative drag reduction technologies is crucial for reducing the environmental impact of the shipping industry and improving its competitiveness.
One such technology is air lubrication, which implements the conceptually simple idea of introducing air \rev{between the water and the ship hull in order to reduce the total area of the wetted surface}.
Works on air lubrication date as early as the 1960s~\citep{Butuzov1966, Butuzov1990}, and planing boats employing the technology have been serially produced~\citep{Matveev2007}.
For large displacement vessels, the adoption has generally gone at a slower pace, yet several commercial air lubrication systems exist, multiple full scale sea trials have been conducted, and at least 23 vessels have an air lubrication system installed on board to date~\citep{ABS2019}.

\begin{figure}[htp!]
	\centering
	\includegraphics[]{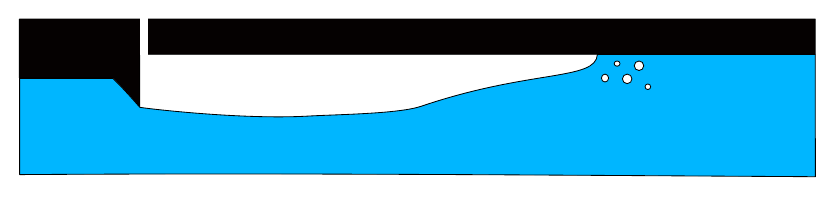}
	\includegraphics[]{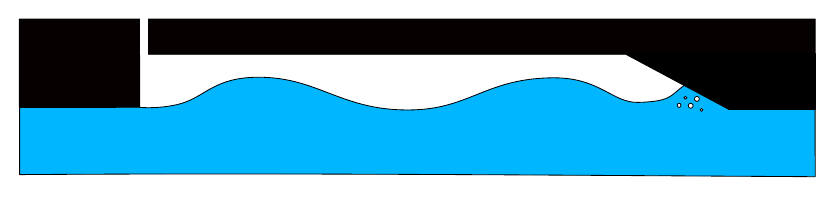}

	\caption
	{
		Sketch of an external (\textit{top}) and internal (\textit{bottom}) air cavity.
	}
	\label{fig:cavity_sketch}
\end{figure}

Three air lubrication techniques are typically distinguished~\citep{Makiharju2012, Zverkhovskyi2014, ABS2019}.
The first is bubble drag reduction, in which, as the name implies, the injected air is mixed with water, surrounding (a part of) the hull with bubbles.
As the flux of air is increased, it is possible that it forms a continuous layer attached to the hull.
This is referred to as air layer drag reduction.
The transition between these two regimes has been studied in~\citep{Elbing2008}.
Finally, the third technique is air cavity drag reduction, which can be implemented in two ways.
One approach, referred to as the external air cavity, is to introduce a spanwise obstruction (a cavitator), behind which the cavity will be formed.
In some works, this is considered to be a variant of the layer drag reduction technique~\citep{Butterworth2015}.
Alternatively, a recess in the hull can be made, for the purpose of better trapping the air.
This technique is referred to as the internal air cavity or the air chamber.
The inclined wall to which the cavity is attached downstream is commonly called the `beach'.
A sketch illustrating both air cavity approaches is given in Figure~\ref{fig:cavity_sketch}.
Typically, several cavities are necessary to cover a significant portion of the hull.

\rev{In any type of air lubrication system, the energy savings due to the reduced drag have to outweigh the additional consumption associated with maintaining the air under the hull.
For the air layer and cavity approaches it is therefore necessary to minimize the risk of leakage and disruption of the attached air.
Arguably, an internal air cavity makes this goal easier to achieve as the air is surrounded by walls from all sides.
In~\citep{Makiharju2012}, the energy gain for air layer and air cavity types of lubrication is estimated to be 10-20\%.
In the case of bubble drag reduction, the challenge is not to allow the released bubbles to escape from under the hull.
In certain cases, hull modifications, such as side-plates, may be necessary to achieve this~\citep{Hoang2009}.
Results from full-scale sea trials of large displacement vessels equipped with bubble drag reduction systems can be found in~\citep{ABS2019}.
The reported net energy savings fall into the range of 2\% to 10\%.}
A more detailed overview of different types of air lubrication methods can be found in~\citep{Ceccio2010, Zverkhovskyi2014, ABS2019}.
The focus of this work is on internal air cavities.
In what follows, selected works on this air lubrication technique are briefly reviewed.

The majority of the published studies report results from physical experiments.
Perhaps the most fundamental ones are~\citep{Lay2010} and~\citep{Makiharju2010}, in both of which the experiments were performed at the U.S. Navy’s W. B. Morgan Large Cavitation Channel.
In~\citep{Lay2010}, a 17.8 cm deep cavity installed on a 12 m long flat plate is tested.
Fluxes necessary to establish and maintain a stable cavity are reported for different free-stream velocities.
The stability of the cavity is analysed with respect to changes in the air influx and perturbations in the free-stream velocity and pressure.
The conclusion is that the cavity is stable with respect to both, under the condition that the changes in the free-stream parameters are gradual.
For a range of conditions, shedding of bubble sheets is reported in the closure region, qualitatively similar to sheet-cloud cavitation on hydrofoils.
In~\citep{Makiharju2010}, the cavity formation is analysed across several Froude numbers, beach wall configurations and spanwise tilts.
Further, the effect of large-scale flow perturbations mimicking ambient waves is considered.
The air flux necessary to maintain the cavity is shown to increase with the size of the perturbations.
Under steady conditions, air shedding occurred in the closure region, presumably due to re-entrant jets and `wave pinch-off'.
Under transient conditions, the behaviour in the closure region changed with the phase of the perturbation, with different patterns of air shedding occurring.

In~\citep{Shiri2012}, cavitation tunnel experiments on a 6 m long cavity with variable height are reported.
Different cavity pressures, cavity heights, and beach wall angles are considered.
The forces on the cavity walls as well as on the aft flat plate are reported for each considered flow condition.
The total force is shown to decrease with increased cavity pressure, due to a smaller amplitude of the cavity wave resulting in a smaller wetted area of the beach wall.
On the other hand, increased pressure also led to higher leakage and thus a higher air flux necessary to maintain the cavity.

Studies of air lubrication using numerical simulations are also present in the literature.
Several works employ potential flow theory, including the pioneering studies of Butuzov~\citep{Butuzov1966, Butuzov1988}.
One approach is to apply this framework in conjunction with other assumptions, which allows deriving approximate expressions for certain properties of the cavity that can be very useful in the initial phase of design.
Of particular relevance for this work is the study by Matveev regarding wave properties of internal air cavities~\citep{Matveev2007}, but see also~\citep{Matveev2003, Li2008}.
Potential flow theory can also be used as basis for purely numerical predictions, see~\citep{Matveev2009, Matveev2012, Matveev2015}.
However, in order to model complicated physical phenomena, such as air leakage, higher-fidelity modelling is required.

As a step towards this, in a range of studies air lubrication is simulated by solving the Reynolds-averaged Navier-Stokes (RANS) equations in conjunction with the Volume of Fluid method for capturing the air-water interface.
In~\citep{Shiri2012}, simulation results complementing the experimental work (discussed above) are reported, but the analysis of the flow is kept at a rather superficial level.
A notable outcome is that two-dimensional simulations failed to yield useful results.
The same case is considered in~\citep{Menon2016}, where the focus is on numerics, in particular, comparing the accuracy of different schemes for interface capturing.
In~\citep{Rotte2018a, Rotte2020}, the possibilities for improving the accuracy of RANS models near the air-water interface are explored.
For the case of an external cavity, good agreement with experimental data is obtained for the velocity profiles.
However, only limited simulation accuracy could be achieved in the closure region, for both the internal and external cavity flows.
It is also suggested that the leakage mechanisms for the two cavity types may be different.
A hybridization of RANS with a scale-resolving turbulence modelling method is considered in~\citep{Rotte2018} to explore the interaction of the detached inflow turbulent boundary layer with the cavity.
Finally, in~\citep{Cucinotta2018, Hao2019} real air-lubricated ship hulls in model scale are simulated.

No study published to date focuses on of the dynamics of the flow near the beach wall.
Yet understanding the flow's behaviour in this region is crucial for minimizing air leakage and improving the stability of the cavity.
As reviewed above, some possible air leakage mechanisms have been proposed, but it is not yet clear to what extent these suggestions are conclusive.
Filling this gap in the literature is the goal of the current work.
To that end, a large-eddy simulation (LES) of the flow in a cavity with the same geometry as in~\citep{Shiri2012} was conducted.
It is noted that this is the first application of this scale-resolving turbulence modelling approach to this type of flows.
LES involves a significantly smaller degree of modelling compared to RANS and offers the possibility to resolve the interaction between the air-water interface and turbulence more accurately.
The reported results include the distributions of statistical moments of pressure, velocity, and the volume fraction of water, and also spatial two-point correlations and temporal energy spectra of velocity.
Together they form a holistic picture of the flow's behaviour and elucidate the dynamics of air leakage, which turn out to differ from what has been suggested previously.
In particular, it is shown that the air entrainment is driven by recirculating flow taking place in the closure region, and the recirculation itself is due to the pressure gradient present at the beach wall.

This article is accompanied by a publicly-available dataset\footnote{DOI: 10.6084/m9.figshare.11698602}. It includes ready-to-run simulation cases, all the obtained results (three- and two-dimensional solution fields, gathered probe data, etc), and also Python scripts that reproduce the figures from the paper.

The remainder of the paper is structured as follows.
The employed methods of computational fluid dynamics are presented in Section~\ref{sec:cfd}.
The simulation set-up is discussed in Section~\ref{sec:setup}, followed by the presentation and analysis of the results in Section~\ref{sec:results}.
Concluding remarks are given in Section~\ref{sec:conclusions}.

\section{Computational fluid dynamics methods} \label{sec:cfd}
This section presents the mathematical model for the flow of two incompressible immiscible fluids used in this paper, as well as numerical methods that are used to solve the corresponding governing equations.

\subsection{Governing equations}
To model the flow, the Volume of Fluid (VoF) method~\citep{Hirt1981} is used.
In VoF, a single set of governing equations is written for the whole flow field, whereas the two fluids are distinguished based on the local value of $\alpha$, which denotes the cell volume fraction occupied by one of the phases, here water.  An additional transport equation for $\alpha$ is solved.
A detailed presentation of the here-employed VoF approach in the context of finite volume solvers can be found in~\citep{Rusche2002, Damian2013}.
What follows is a concise summary.

The momentum and continuity equations read as follows,
\begin{align}
\pdiff{\rho u_i}{t}+\pdiff{}{x_j}\left(\rho u_iu_j \right) & = -\pdiff{p_{\rho gh}}{x_i} - g_ix_i\pdiff{\rho}{x_i} +  \pdiff{}{x_j}\left( \mu \left( \pdiff{u_i}{x_j} + \pdiff{u_j}{x_i} \right) \right) ,\quad (i=1,2,3) \label{eq:lesmom} \\
\label{eq:lescont}
\pdiff{u_j}{x_j} & =  0.
\end{align}
Here, summation is implied for repeated indices, $u_i$ is the velocity, $\rho$ is the density, $\mu$ is the dynamic viscosity,  $g_i$ is the standard acceleration due to gravity, and $p_{\rho gh} = p - \rho g_3x_3$ is the dynamic pressure.

The transport equation for the volume fraction $\alpha$ reads as
\begin{equation}
\label{eq:alpha}
	\pdiff{\alpha}{x_i} + \pdiff{u_j \alpha }{x_j} + \pdiff{}{x_j}\left(u^r_j (1 -\alpha)\alpha\right) = 0.
\end{equation}
The last term in the equation is artificial and its purpose is to introduce additional compression of the interface.
To that end, the direction of $u_i^r$ is aligned with the interface normal, $n_i^f$, which is computed as
\begin{equation} \label{eq:normal}
n^f_i= \pdiff{\alpha}{x_i} \bigg/ \left( \bigg\vert \pdiff{\alpha}{x_i}  \bigg\vert + \delta_N \right) .
\end{equation}
Here, $\delta_N$ is a small number added for the sake of numerical stability.
The magnitude of $u^r_i$ is defined as $C_\alpha |u_i|$, where $C_\alpha = 1$ is an adjustable constant.

It is noted that including surface tension effects via the Continuous Force Model has been attempted, but resulted in numerical instabilities that increased in severity with the increase in the density of the computational mesh.
Consequently, the main simulation results presented below are obtained with the surface tension effects ignored.
However, in Appendix~\ref{app:tension} a comparison between results obtained with and without surface tension modelling is presented, with the corresponding simulations run on a coarser grid, thus avoiding problems with instability.
\rev{As the main results are not significantly changed on this coarser mesh, it is concluded that surface tension does not have a significant effect on the flow.}

Given the value of $\alpha$, the local material properties of the fluid are computed as
\begin{align}
& \rho = \alpha\rho_{w} + (1 - \alpha)\rho_{a},
& \mu = \alpha\mu_{w} + (1 - \alpha)\mu_{a},
\end{align}
where $\rho_{w} = 999$ $\mathrm{kg/m^3}$,  $\rho_{a} = 2.5$ $\mathrm{kg/m^3}$,
$\mu_{w} =  1.14 \cdot 10^{-3}$ $\mathrm{Pa \cdot s}$, and \mbox{$\mu_{a} = 1.48 \cdot 10^{-5}$ $\mathrm{Pa \cdot s}$}, with the $w$ and $a$ subscripts denoting water and air respectively.
\rev{The increased density of air is motivated physically by the fact that the vessel with the cavity onboard can be expected to have a draft of at least 20 m.
However, higher air density also greatly benefits numerical stability, which has generally been difficult to achieve.}
Unfortunately, it also \rev{limits} direct comparison with the experimental results reported in~\citep{Shiri2012} for the same geometry and otherwise similar flow conditions.

In order to obtain the governing equations for LES based on the above two-phase flow model, spatial filtering should be applied to~\eqref{eq:lesmom}-\eqref{eq:alpha}.
In practice, explicit filtering is often not used due to the fact that in the discrete setting the density of the numerical grid implicitly defines the length scale of the smallest eddies that can be resolved.
In this work, the finite volume method (FVM) is used for discretization, as discussed in more detail in the next subsection.
In the case of FVM, the width of the implicit filter is defined as the cubic root of the local computational cell volume.

A consequence of filtering the equations is the appearance of the subgrid stress term in the momentum equation~\eqref{eq:lesmom}, which has to be modelled in order to close the system of equations.
A variety of closures have been proposed, the majority of which (see e.g.~\citep{Smagorinsky1963, Yoshizawa1982, Kim1995, Nicoud1999a}) are based on the Boussinesq approximation, i.e.~that the subgrid stresses have the same structure as the viscous ones.
In this work, no such closure is employed and the subgrid stress term is dropped from the equations.
\rev{This decision is chiefly based on the choice of numerical schemes, which are discussed in next subsection.
However, an additional factor is that the existing closures were developed and validated for single phase flows, and their predictive accuracy for multiphase flows is not well documented.}

\subsection{Numerical methods} \label{sec:numerics}

The LES was performed using the open-source computational fluid dynamics software OpenFOAM\textsuperscript{\textregistered} version 1806.
In particular, the solver \ttt{interFoam}, distributed with this code, was employed.
OpenFOAM\textsuperscript{\textregistered} uses a cell-centred FVM approach for discretising the governing equations.
Details about this method and its application to fluid dynamics can be found in textbooks~\citep{Ferziger2002, Versteeg2007} and OpenFOAM\textsuperscript{\textregistered}-related theses, for example~\citep{Jasak1996, Rusche2002, Damian2013}.
In brief, the computational domain is decomposed into convex polyhedral cells, and an integral form of the governing equations, valid for each cell volume, is derived.
Where necessary, the Gauss-Ostrogradsky theorem is used to convert volume integrals into integrals across the surface of the cell, which are in turn approximated by sums across the cell faces.
To evaluate the sums, the values of the unknowns at the face centres are required, and a crucial step of the numerical procedure is obtaining them given the values at the centroids of the cells and a suitable interpolation scheme.

The most trivial choice is using linear interpolation, which is second-order accurate.
This scheme can be applied to interpolation of diffusive fluxes without negative side-effects.
Unfortunately, this does apply to convective fluxes, and in this case linear interpolation introduces parasitic oscillations.
Nevertheless, linearly interpolating the convective fluxes is a common practice in LES studies of canonical single-phase flows, because the density of the mesh can be kept high enough for the introduced errors to be negligible.
In this study, such a dense grid could not be afforded and additionally it is expected that near the interface even small oscillations could lead to significant stability issues.
Therefore, a second-order accurate upwind scheme is employed instead.
Although this scheme is also unbounded, the parasitic oscillations are damped by numerical dissipation.

A time integration scheme should also be chosen, and for \texttt{interFoam} the available choices are a first-order accurate implicit Euler scheme, a second-order accurate semi-implicit Crank-Nicholson scheme, and a linear blending of the two.
Somewhat surprisingly, a linear blending with a 90\% weight assigned to the Crank-Nicholson scheme could be used during initial testing with simulations on coarser meshes, but on the densest mesh, which was used to compute the final results, the simulation could only be stabilized using the Euler scheme.

In summary, the overall level of dissipativity of the numerics used in the simulation is higher than that classically used for LES.
This is not unusual for simulations of industrial flow problems, in which one typically has to settle for a compromise between accuracy and stability.
\rev{However, careful analysis of the results is necessary to confirm that a sufficiently broad range of turbulent scales is resolved in the simulation.
To that end, energy spectra and two-point autocorrelation functions of the velocity field are examined in Section~\ref{sec:results}, and it is demonstrated that the conducted simulation classifies as an LES.}

\rev{Returning to the question of subgrid scale modelling, in the experience of the authors, classical Boussinesq-type closures do not interact well with the chosen set of numerical schemes.
The issue is that the introduced numerical dissipation is comparable in magnitude with that produced by the model.
More accurate results are thus achieved by simply ignoring the subgrid stresses.}
This approach is sometimes referred to as implicit (I)LES but this is a rather loose application of this term because for proper ILES the leading truncation error term of the numerical schemes should be tailored to have a structure similar to that of the subgrid stresses~\citep{Grinstein2007}.

\section{Simulation set-up} \label{sec:setup}
\begin{figure}[htp!]
	\centering
	\includegraphics{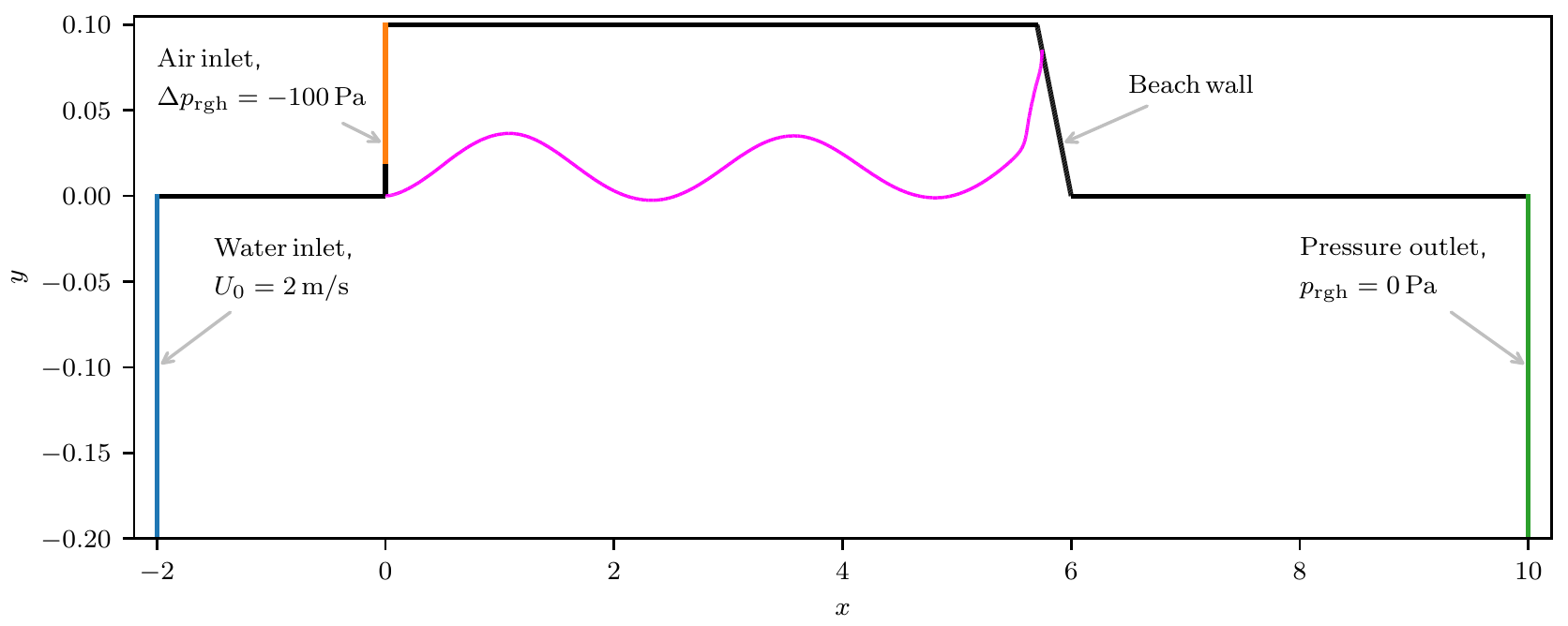}
	\caption
	{Set-up of the simulation. The bottom wall of the domain, not shown in the figure, is located at $y = -1.5$ m. The magenta line shows the $\mean{\alpha} = 0.5$ isoline.}
	\label{fig:case}
\end{figure}

The simplified hull cavity profile used in this study has previously been employed in a series of experiments at the SSPA cavitation tunnel in Sweden~\citep{Shiri2012}.
A side-view of the computational domain is shown in Figure~\ref{fig:case}.
The considered cavity has length $L_c = 6$ m and height $H_c = 0.1$ m, and the slope of the beach wall is 1:3, corresponding to an inclination angle of 18.4 degrees.
The left side of the cavity is split between a trim wall reaching $y = 0.02$ m and the air inlet, where a total pressure\footnote{That is, $p_{\rho gh} + 0.5\rho u_iu_i$. Not to be confused with the sum of the static and dynamic pressures $\rho g_i x_i + p_{\rho gh} = p$.} of $-100$~Pa is prescribed.
Thus, here the cavity pressure is prescribed, whereas the flux of air through the inlet is allowed to be transient.
Using the total pressure as opposed to just prescribing $p_{\rho gh}$ leads to better numerical stability in the initial phase of the simulation, and the velocity at the inlet is so small that physically the difference is negligible.
At the outlet on the right side of the domain, $p_\text{rgh} = 0$~Pa is enforced, which leads to the nominal pressure difference between the phases being $\Delta p_\text{rgh} = - 100$~Pa.
At the water inlet, located 2 m upstream of the cavity, a uniform velocity $U_0 = 2$~m/s is prescribed.
In practice, a turbulent boundary layer can be expected to approach the cavity instead, but no significant interactions between the inflow turbulence and the flow in the cavity are reported in~\citep{Shiri2012}.
The size of the domain in the spanwise direction is $L_z = 0.25$~m, with periodic boundary conditions applied on the associated boundaries.
This avoids any effects of the side-walls on the wave system inside the cavity, the study of which is outside of the scope of this work.
The particular choice of the value of $L_z$ is justified below based on analysis of spanwise two-point correlations of the unknowns.

Based on the above, it is possible to define the Reynolds, Froude and cavity pressure difference numbers of the flow
\begin{align}
& Re = \frac{\rho_w U_0 L_c}{\mu_w} \approx 1.05 \cdot 10^7, \\
& Fn = \frac{U_0}{\sqrt{gL_c}} \approx 0.26, \\
& \sigma = \frac{\Delta p_{\rho gh}}{0.5\rho_w U^2_0} \approx 0.05.
\end{align}

As discussed in detail below, the flow in the cavity can roughly be divided into two regions.
The first~$\approx 5.4$~m are occupied by a two-dimensional steady gravity wave, see the magenta line in Figure~\ref{fig:case}.
By contrast, in the closure region the flow becomes highly unsteady and turbulent.
Consequently, a computationally efficient mesh should be dense and isotropic in the closure region, and highly anisotropic in the region of the steady wave, where the resolution in the vertical direction should be much larger than in the streamwise, and spanwise resolution is redundant.
To accommodate the change in the spanwise resolution, an unstructured grid is required.
However, LES is commonly performed on structured hexahedral meshes since they benefit the accuracy of numerical interpolation schemes.

Here, a special mesh construction approach using adaptive mesh refinement (AMR) is employed.
The AMR can be used to refine selected hexahedral cells by dividing them along all three spatial directions, thus producing eight cells in place of one.
Which cells are to be refined is controlled based on the values of some user-defined field, $R_\text{amr}$.
Another input parameter is the amount of times refinement can be applied to a particular cell, which is referred to as the refinement level.
It is stressed that in OpenFOAM no special numerical treatment is required for handling the fluxes between a coarser cell and two neighbouring refined cells.
The face of the coarse cell is simply split into two, meaning that the cell is actually no longer a hexahedron, but a polyhedron with two connected faces parallel to each other.

First, a coarse structured hexahedral mesh was created, with isotropic cells in the closure region and cells elongated in the streamwise direction in the steady wave region and the rest of the domain.
Some refinement in the vertical direction was also introduced where the wave was expected to be located (based on linear theory and test simulations).
This initial mesh was then subject to AMR based on the following definition of $R_\text{amr}$:
\begin{equation*}
R_\text{amr} =
\begin{cases}
\alpha & x \leq 5.4 \\
0.5 & x \in (5.4, 6.05), \; y > -0.01 \\
0 &  \text{otherwise.}
\end{cases}
\end{equation*}
Refinement was prescribed for cells in which $R_\text{amr} \in [0.02, 0.98]$, and the refinement level set to 3.
As a result, in the steady wave region the refinement follows the location of the interface and in the closure region the mesh is always kept at the maximum level of refinement, producing a mesh suitable for LES.
In the rest of the domain, no refinement is applied leading to significant savings in computational resources.
Note that the mesh is dynamic only in the initial phase of the simulation, during which the profile of the wave is formed.
Once the wave is steady, the mesh ceases to change, with its size being $\approx 25 \cdot 10^6$ cells.

\begin{figure}[htp!]
	\centering
	\includegraphics[width=0.75\linewidth]{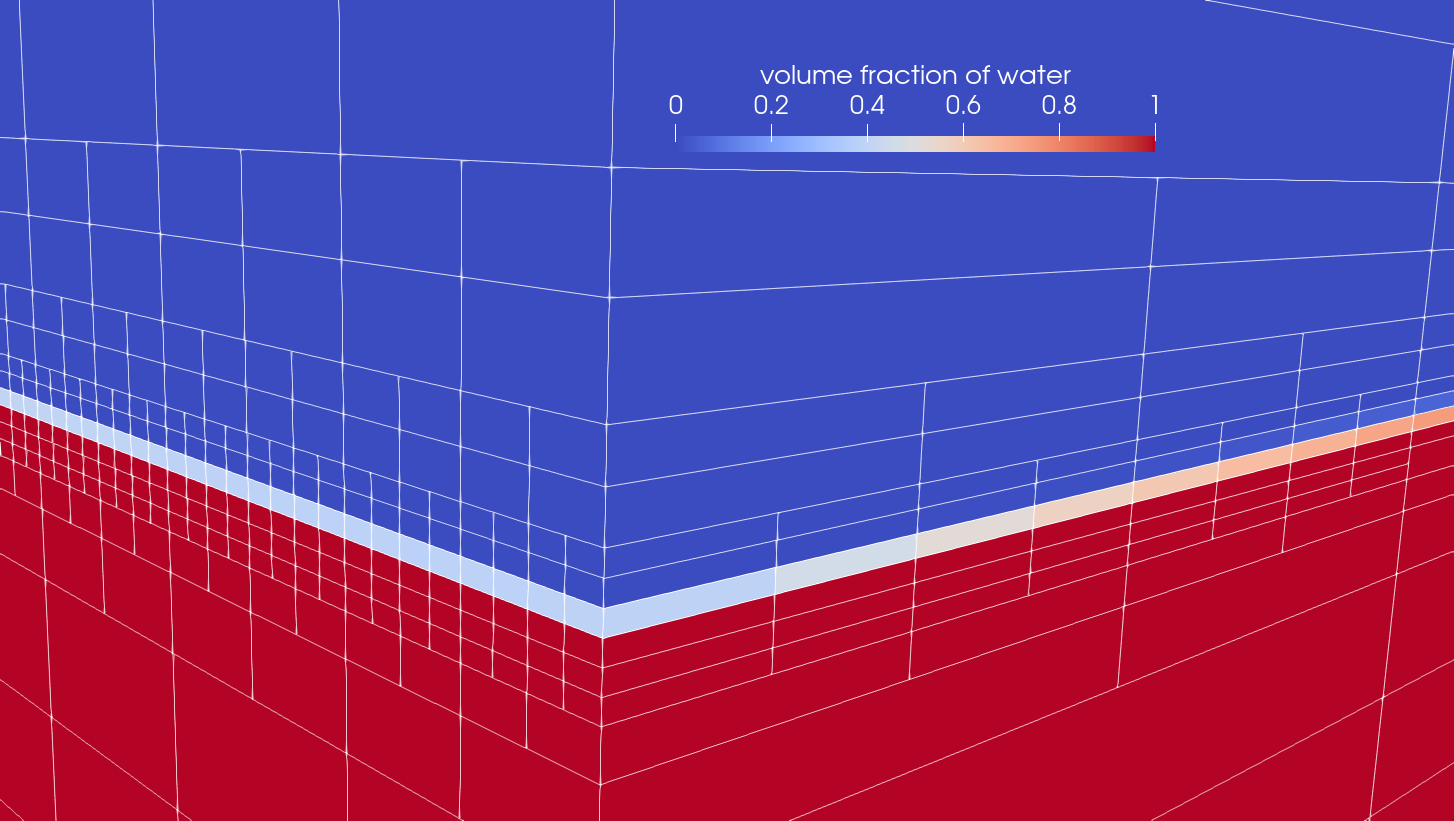}
	\caption
	{
		A spanwise cut through the domain exposing the mesh structure in the region occupied by the steady wave.
		A close-up on the interface.
	}
	\label{fig:mesh}
\end{figure}

In Figure~\ref{fig:mesh}, a close-up of the mesh at the interface of the steady wave is shown.
A cut in the spanwise direction is made to expose its three-dimensional structure.
As expected, the most refined cells are eight time smaller than the initial cell size in each spatial direction.
Unfortunately, since the AMR divides the cells in the spanwise direction as well, some redundant resolution could not be avoided, but since it is only present near the interface the savings compared to a fully structured mesh are large.

An analysis of the resolution of the mesh with respect to the size of the turbulent structures in the flow is presented in the next section.
Additionally, results from simulations on meshes constructed by setting the refinement level to, respectively, 0 (no AMR), 1, and 2, are presented in Appendix~\ref{app:mesh} and compared to the outcomes of the main simulation.
It is demonstrated that the simulation using the refinement level 2 mesh produces distributions of statistical moments of the flow variables that are very close to those from the main simulation.
Further refining the mesh is therefore not expected to give significant gains in predictive accuracy.

To keep the simulation stable, a small time-step had to be employed, $\Delta t = 2 \cdot 10^{-5}$ s.
This allowed to keep the maximum Courant number in the domain below 0.35 during the whole simulation.
As initial conditions, a flow-field obtained from a previous computation using a coarser mesh and a slightly different value for the air density was employed.\footnote{That simulation had, in turn, been initialized with uniform velocity and pressure values and a flat air-water interface.}
A total of 25 s of simulation time were dedicated to removing any artificial transient behaviour associated with initial conditions.
This amount is likely to be somewhat excessive.
However, interpolation from a coarser mesh and the AMR introduce artefacts into the solution, thus it was crucial to make sure that a steady wave profile had been reached before the gathering of statistical data commenced.
A period of 14.5 s was dedicated to the latter.

\section{Results} \label{sec:results}
In this section, the results of the conducted LES are presented and discussed.
The properties of the steady wave in the cavity are considered first.
In particular, it is interesting to see how well they can be predicted based on linearised flow theory.
To make comparisons, the location of the air-water interface should first be formally defined based on the values of the flow field.
In VoF, the common approach is to define it as the $\alpha = 0.5$ isosurface, which will be denoted as $\alpha_{0.5}$.
\rev{The mean location of the interface, $\mean{\alpha_{0.5}}$, is obtained as the $\mean{\alpha} = 0.5$ isoline, where the angular brackets denote the temporal and spanwise average.}

\begin{figure}[h!]
	\centering
	\includegraphics[center]{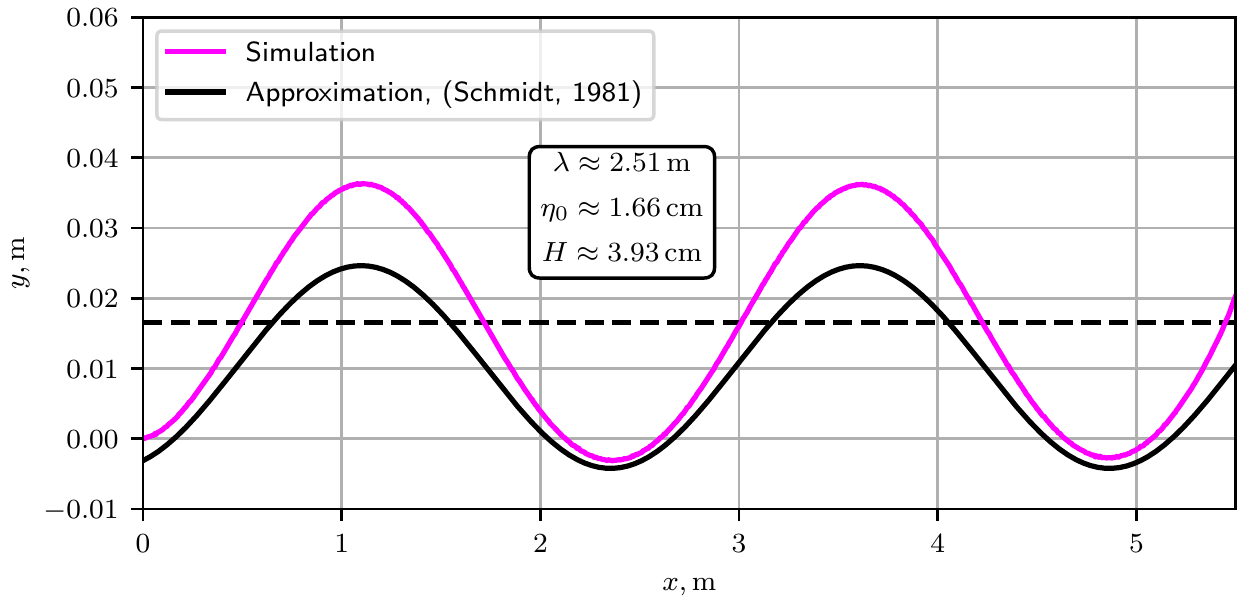}
	\caption
	{
		The profile of the steady wave obtained in the simulation (magenta line) and the transom wave approximation by Schmidt~\citep{Schmidt1981}. (black line).
	}
	\label{fig:linear}
\end{figure}

Figure~\ref{fig:linear} shows the profile of the steady wave obtained in the simulation and also lists its most important properties.
Linear gravity wave theory provides an approximation of the wavelength as $\lambda^\text{thr} \approx 2.56$~m, which agrees well with the value of $\approx 2.51$~m obtained in the simulation.
The average vertical position of the interface,~$\eta_0$, is approximated by the height of the water column necessary to compensate $\Delta p_{\rho g h}$,~i.e. $\eta_0^\text{thr} = \Delta p_{\rho g h}/(\rho_w g) \approx 1.02$~cm.
The obtained value, $\eta_0 \approx 1.66$~cm is quite significantly higher.
A plausible explanation is that the observed discrepancy is an effect of the flow dynamics in the closure region.
However, in this case it would be reasonable for $\eta_0$ to become closer to $\eta_0^\text{thr}$ if one was to increase the length of the cavity without altering the size of the closure region.
As demonstrated in Appendix~\ref{app:cavity_length}, no such trend is observed.
Another explanation is that the pressure at the outlet, which is used to compute the nominal pressure difference $\Delta p_{\rho g h}$, only provides an approximation of the pressure level in the water phase.
The value $\Delta p_{\rho g h}$ corresponding to the obtained value of $\eta_0$ is $\approx -163$ Pa, giving a 63 Pa difference with what is nominally prescribed, which is significant.
To further explore this issue, simulations at different  $\Delta p_{\rho g h}$ should be conducted in order to observe the trend in its relation to $\eta_0$.
Such an investigation is left to be a part of a future study.

In~\citep{Matveev2007}, the potential flow solution for the waves in an infinitely long internal cavity is found to agree well with an approximate analytical solution derived by Schmidt~\citep{Schmidt1981} for transom waves.
The approximation reads as
\begin{equation} \label{eq:schmidt}
\eta = -\sqrt{2}\eta^\text{thr}_0 \cos\left( \frac{2\pi}{\lambda^\text{thr}}x + \frac{\pi}{8}\right) + \eta^\text{thr}_0,
\end{equation}
where $\eta$ is the location of the interface, which is here approximated by $\mean{\alpha_{0.5}}$, as discussed above.
Due to the discrepancy between $\eta_0^\text{thr}$ and $\eta_0$, Schmidt's approximation becomes rather inaccurate, see the black curve in Figure~\ref{fig:linear}.
However, the phase of the wave is predicted remarkably well.

In summary, it is found that both the length and the phase of the wave can be accurately predicted using linear theory, whereas $\eta_0$ and, consequently, the amplitude of the wave are likely to be under-predicted.
This should be taken into account when considering the height of cavity and the trim wall.

\begin{figure}[h!]
	\centering
	\includegraphics[width=0.9\linewidth]{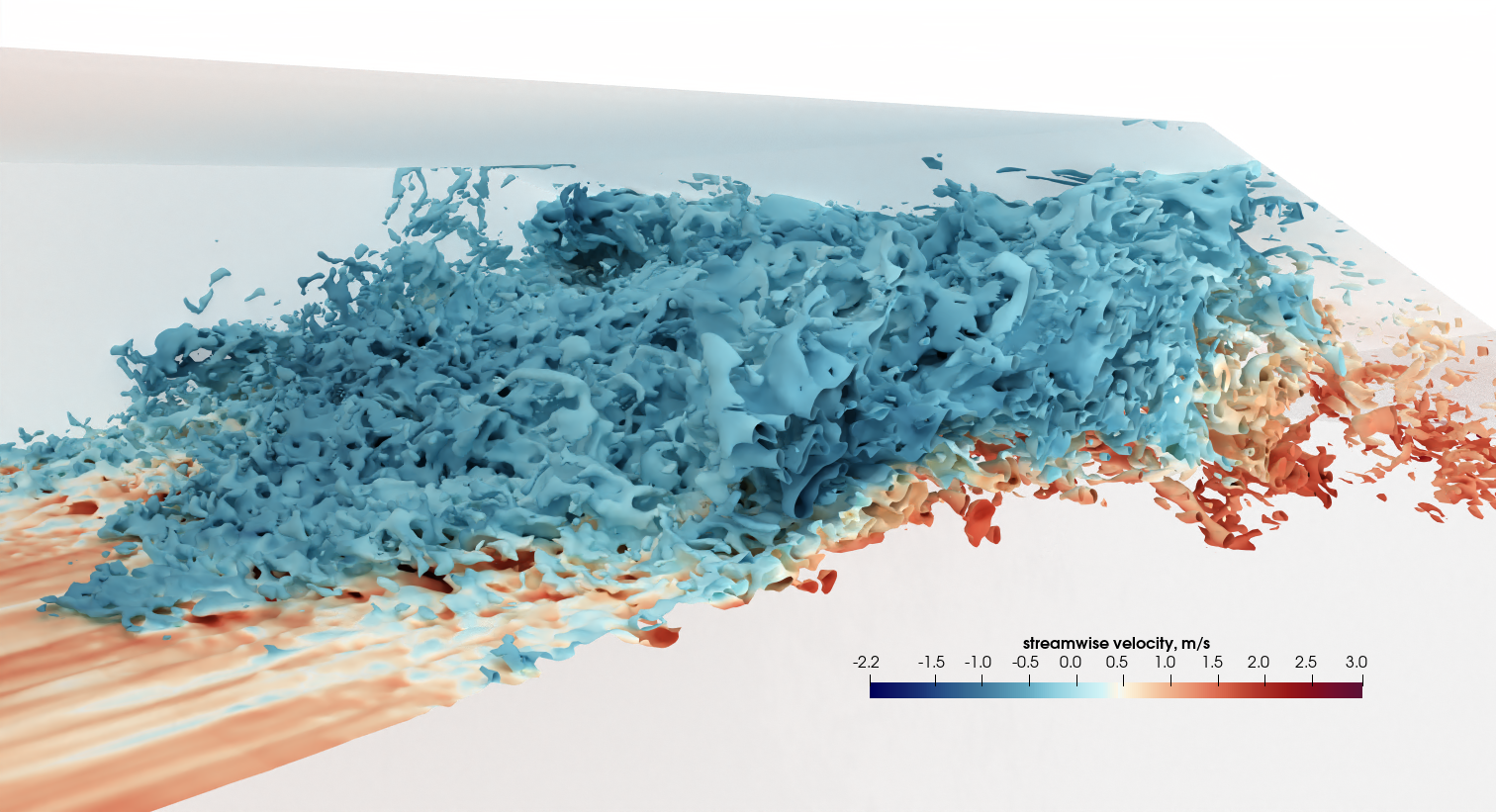}
	\caption
	{
		A snapshot of the air-water interface, $\alpha_{0.5}$, in the closure region, coloured by streamwise velocity.
	}
	\label{fig:alpha_angle}
\end{figure}

Attention is now turned to the characteristics of the flow in the closure region.
First, the snapshot of $\alpha_{0.5}$ shown in Figure~\ref{fig:alpha_angle} is considered.
The plot provides a good qualitative view of the extreme degree of unsteadiness of the flow.
A reverse flow plunging onto the steady wave is observed, and in fact, overturning waves are distinguishable across the whole region.
Violent mixing leads to both detached water droplets and entrainment of air, with swarms of entrained bubbles on their way downstream clearly seen on the right.


\begin{figure}[h!]
	\centering
	\includegraphics[center]{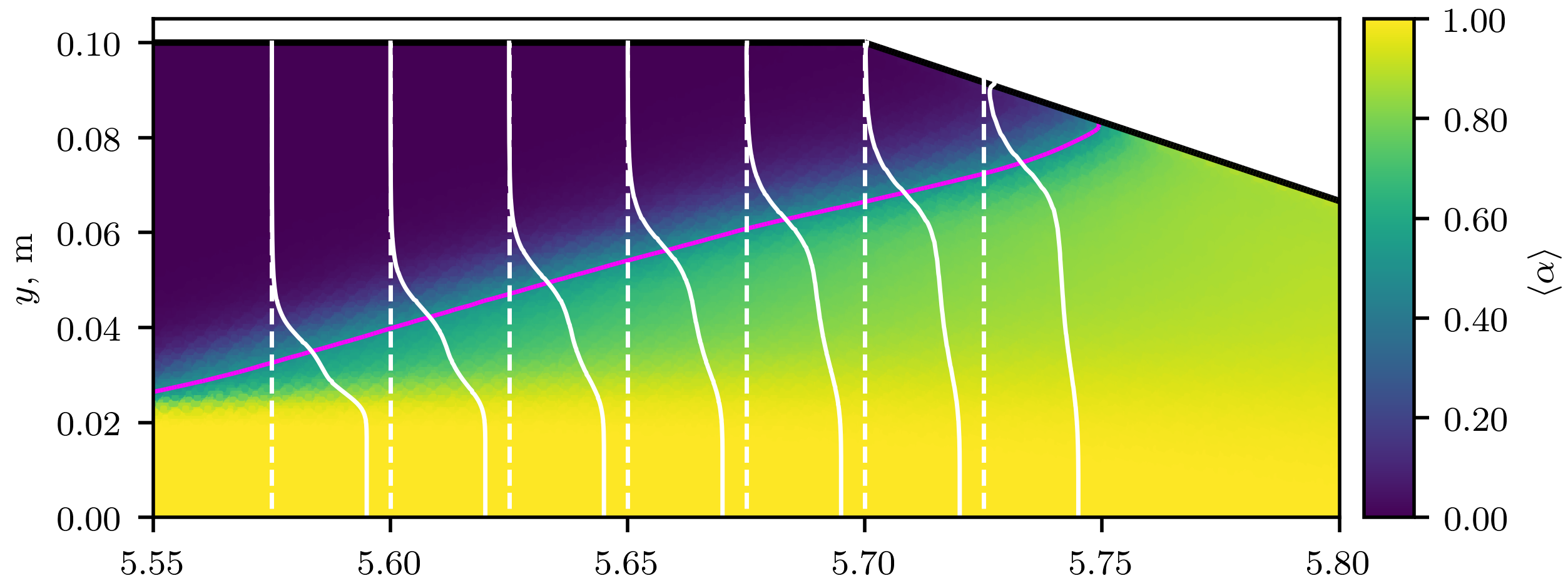}
	\includegraphics[center]{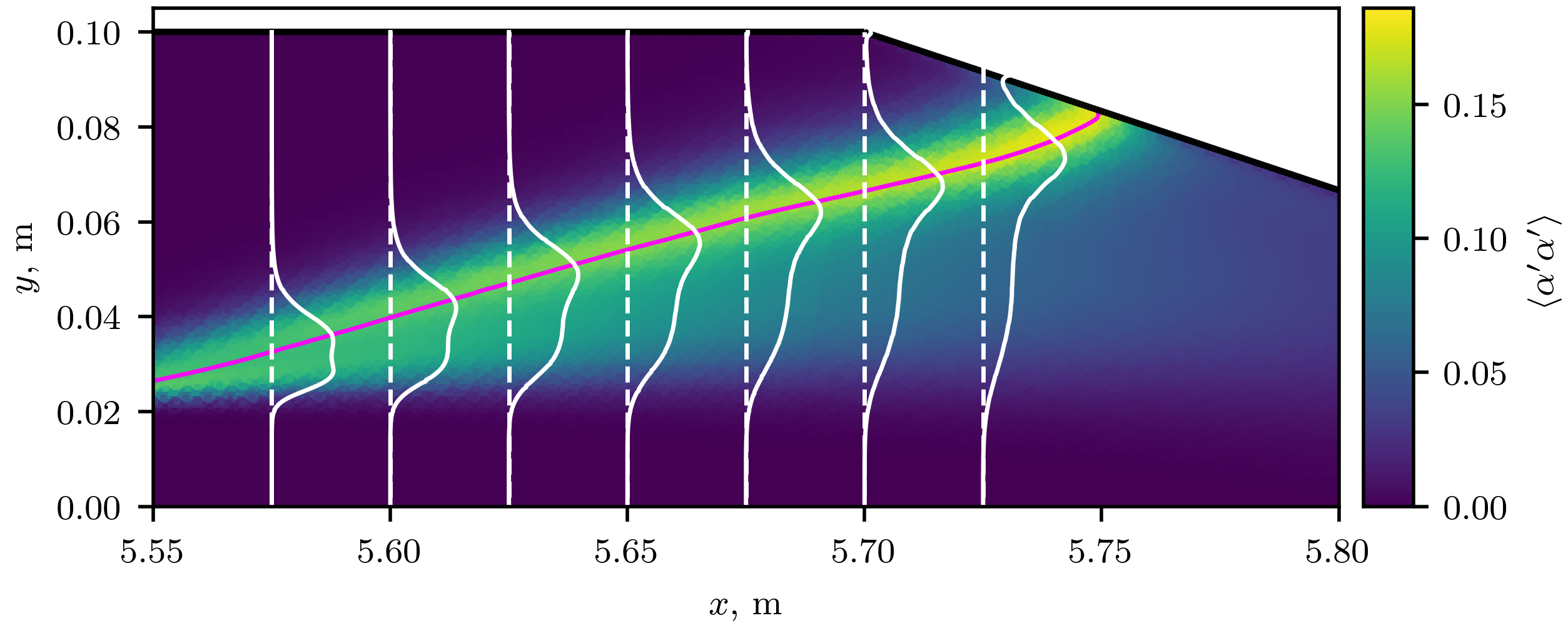}
	\caption
	{
		\textit{Top:} Distribution of the mean volume fraction of water, $\mean{\alpha}$, near the beach wall.
		White solid lines show profiles of $0.02\mean{\alpha}$ at selected locations, marked by dashed white lines.
		The magenta line shows the  $\mean{\alpha_{0.5}}$ isoline.
	}
	\label{fig:alpha_2d}
\end{figure}

The first and second-order statistical moments of $\alpha$ are shown in Figure~\ref{fig:alpha_2d}.
The magenta line shows the $\mean{\alpha_{0.5}}$ isoline.
Confirming the qualitative observations based on Figure~\ref{fig:alpha_angle}, $\mean{\alpha}$ is seen to be less than  $1$ in a significant part of the closure region.
However, the values of $\mean{\alpha}$ also grow noticeably with $x$, for $x \lessapprox 5.76$, implying that only a portion of the entrained air is leaked.
The variance of $\alpha$ indicates that the most intensive mixing occurs near the location of $\mean{\alpha_{0.5}}$, in particular, close to where it attaches to the beach wall.
However, closer to the region where the reverse flow plunges onto the steady wave, the peaks in $\langle \alpha'\alpha'\rangle$ are broader, penetrating deeper under $\mean{\alpha_{0.5}}$.

\begin{figure}[h!]
	\centering
	\includegraphics[]{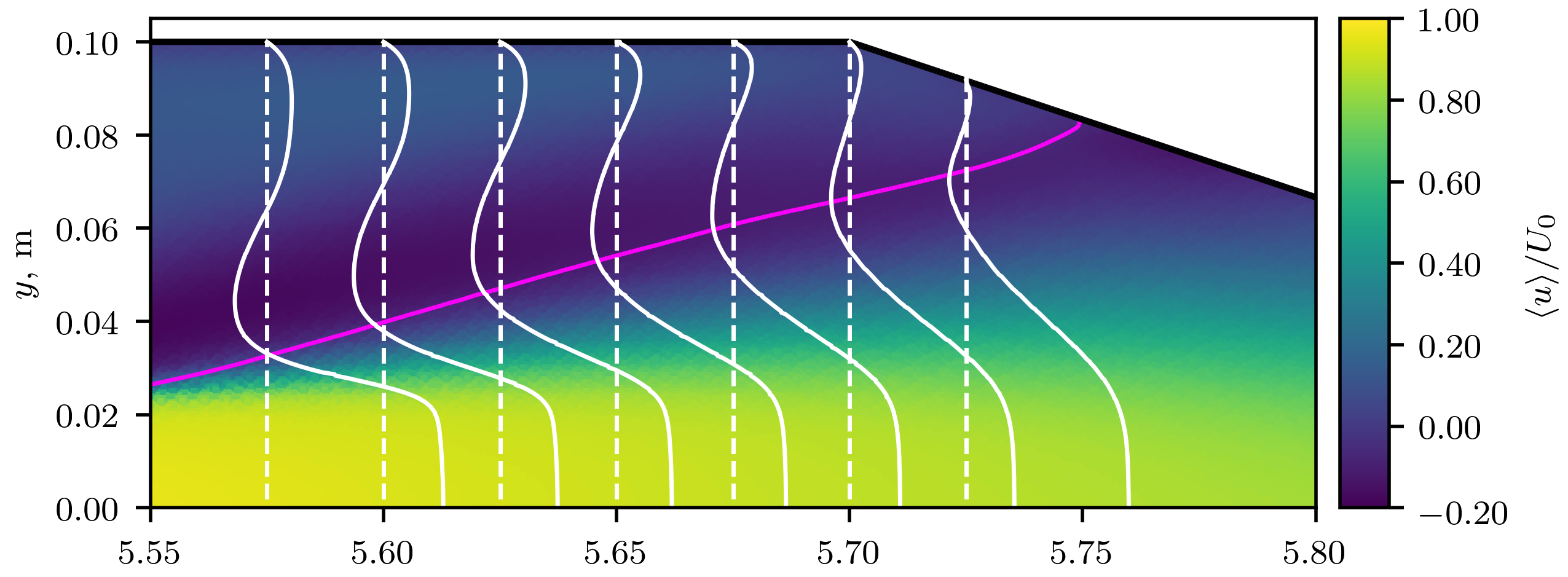}
	\includegraphics[]{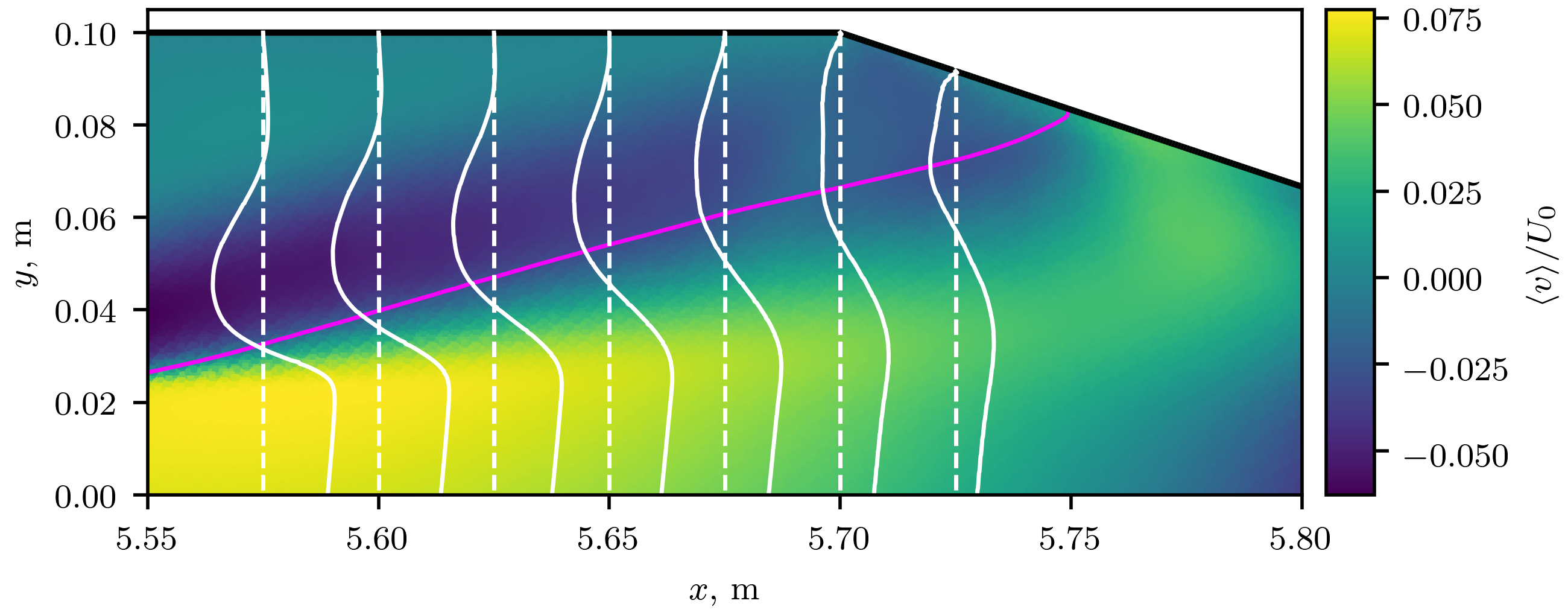}
	\caption
	{
		Distribution of the normalized mean velocity components. \textit{Top}: the streamwise velocity, $\mean{u}/U_0$. \textit{Bottom}: the vertical velocity, $\mean{v}/U_0$.
		White solid lines show profiles of $\mean{u}/(25U_0)$  and $\mean{v}/(5U_0)$ at selected locations on the respective plots, marked by dashed white lines.
	 	The magenta line shows the $\mean{\alpha_{0.5}}$ isoline.
	}
	\label{fig:umean_2d}
\end{figure}

\begin{figure}[htp!]
	\centering
	\includegraphics[center]{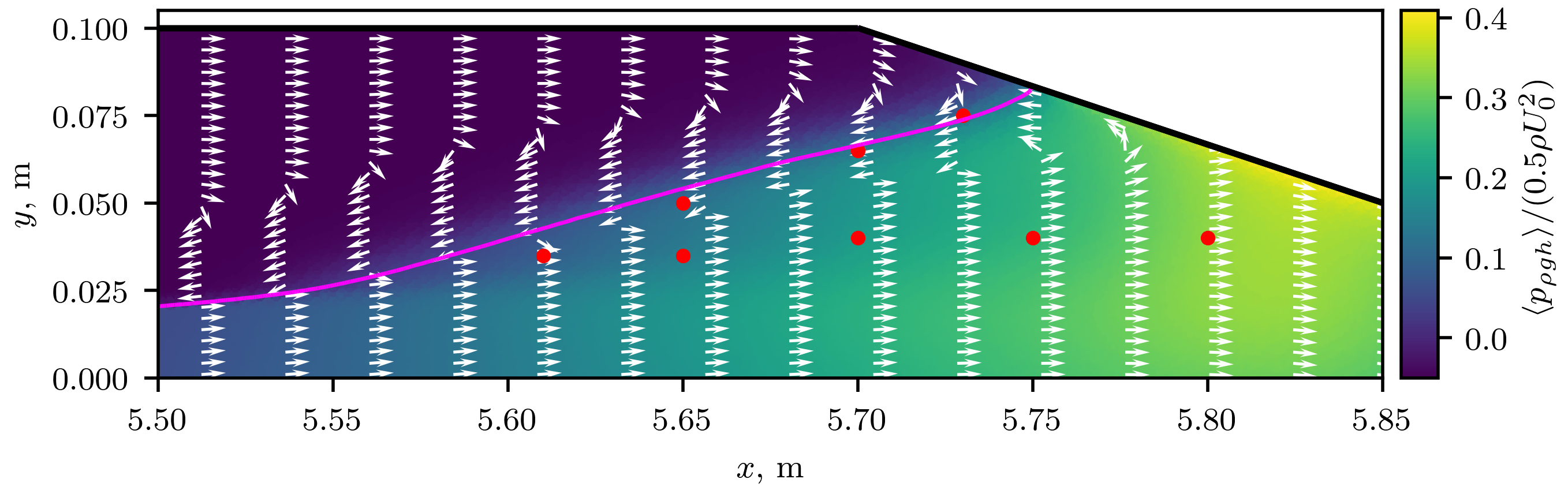}
	\caption
	{
		Distribution of the mean dynamic pressure near the beach wall.
		Arrows indicate the direction of the mean velocity.
		The magenta line shows the  $\mean{\alpha_{0.5}}$ isoline.
		Red dots show locations where data probes have been placed for sampling time-signals of $u$ and $\alpha$.
	}
	\label{fig:recirculation}
\end{figure}

Recall that in Figure~\ref{fig:alpha_angle} it is shown that a significant part of the flow in the closure region is directed upstream with respect to the mean flow.
The same can be observed by looking at the components of the mean velocity field, which are plotted in Figure~\ref{fig:umean_2d}.
In fact, both the streamwise and the vertical components change sign slightly below the mean location of the interface, meaning that it is located in the core of a recirculation zone.
Since this vortex governs the flow dynamics in the closure region, it is critical to understand its origins.
Recall that $\mean{\alpha_{0.5}}$ experiences a steep incline as it approaches the beach wall, see Figure~\ref{fig:case}.
The behaviour of the velocity field indicates that this causes the flow to separate, forming a `detached shear layer'.
Consequently, a recirculation zone is formed above.
In Figure~\ref{fig:recirculation}, it is visualized using a vector plot of the mean velocity field.
This type of separation is well documented for the case of a hydraulic jump (see e.g.~\citep{Mortazavi2016a}), where the recirculation zone is commonly referred to as the `roller region'.
Note that, as is commonly the case, the magnitude of the velocity in the reverse flow is significantly lower than that of the free stream.
However, with decreasing $x$ it grows, which implies that overturning waves plunge with higher force.
This may explain the increased breadth of the $\mean{\alpha'\alpha'}$ peaks discussed above.

To further understand the flow dynamics, the reason for the interface being displaced upwards should be identified.
By design, in an internal air cavity, the water flow stagnates at the beach wall.
Consequently, pressure build-up occurs at the stagnation point, leading to a significant pressure gradient.
This is clearly illustrated in Figure~\ref{fig:recirculation} showing the distribution of $\mean{p_{\rho g h}}$ in the closure region.
The pressure gradient produces a force roughly tangential to the beach wall that pushes the interface upwards.
This effect is unique to internal air cavities and represents an additional complexity factor in their design.

\begin{figure}[h!]
	\centering
	\includegraphics[center]{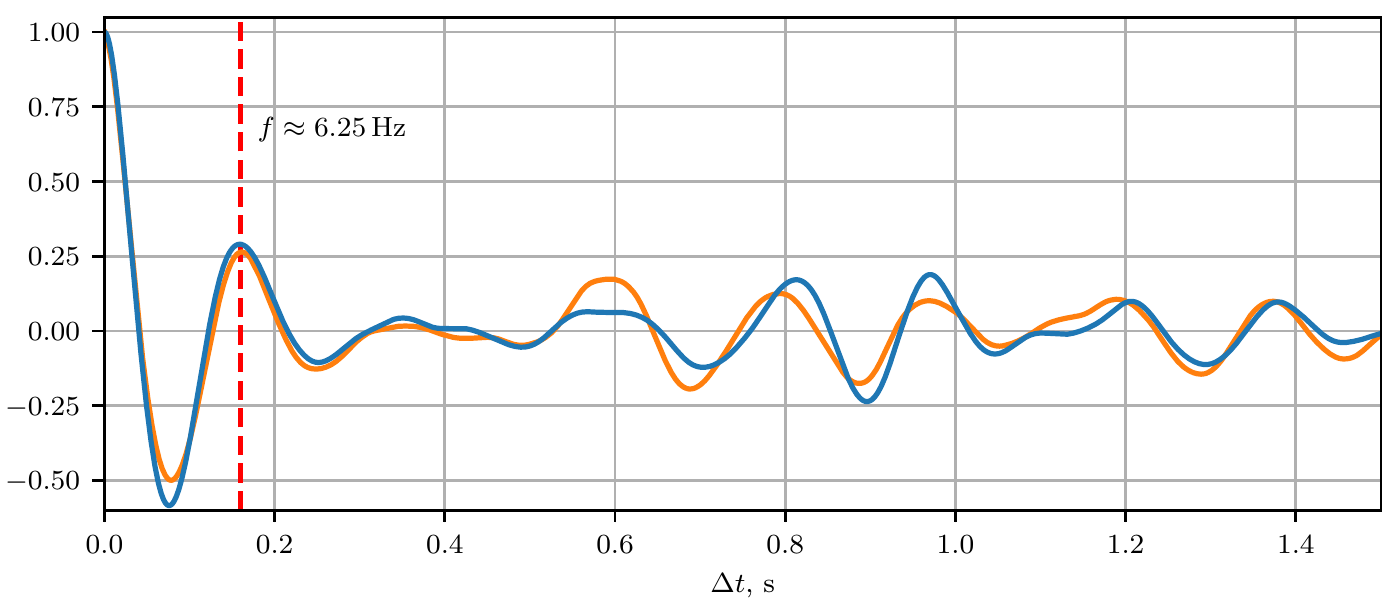}
	\caption
	{
		Blue line: Temporal autocorrelation function of the total air volume present in the box with bounds $x \in [5.8, 5.825]$, $y \in [-0.025, 0.1]$, $z \in [0.0, L_z]$.
		Orange line: Temporal autocorrelation function of the total normal force on the beach wall.
	}
	\label{fig:air_acf}
\end{figure}

Inspection of flow animations suggests that the formation of leaking bubble swarms occurs periodically.
To quantify that, the total volume of air in a box located directly downstream of the reattachemt point of the detached flow ($x \in [5.8, 5.825]$, $y \in [-0.025, 0.1]$, $z \in [0.0, L_z]$) has been computed and stored at each time-step of the simulation.
Additionally, the signal of the total force on the beach wall has also been acquired.
Figure~\ref{fig:air_acf} shows the autocorrelation functions of both signals.
Clearly, they are well-correlated, and both signals exhibit a dominant frequency $f \approx 6.25$~Hz, which corresponds to Strouhal number $St \approx 0.12$, based on $U_0$ and $H$.

\begin{figure}[h!]
	\centering
	\includegraphics[center]{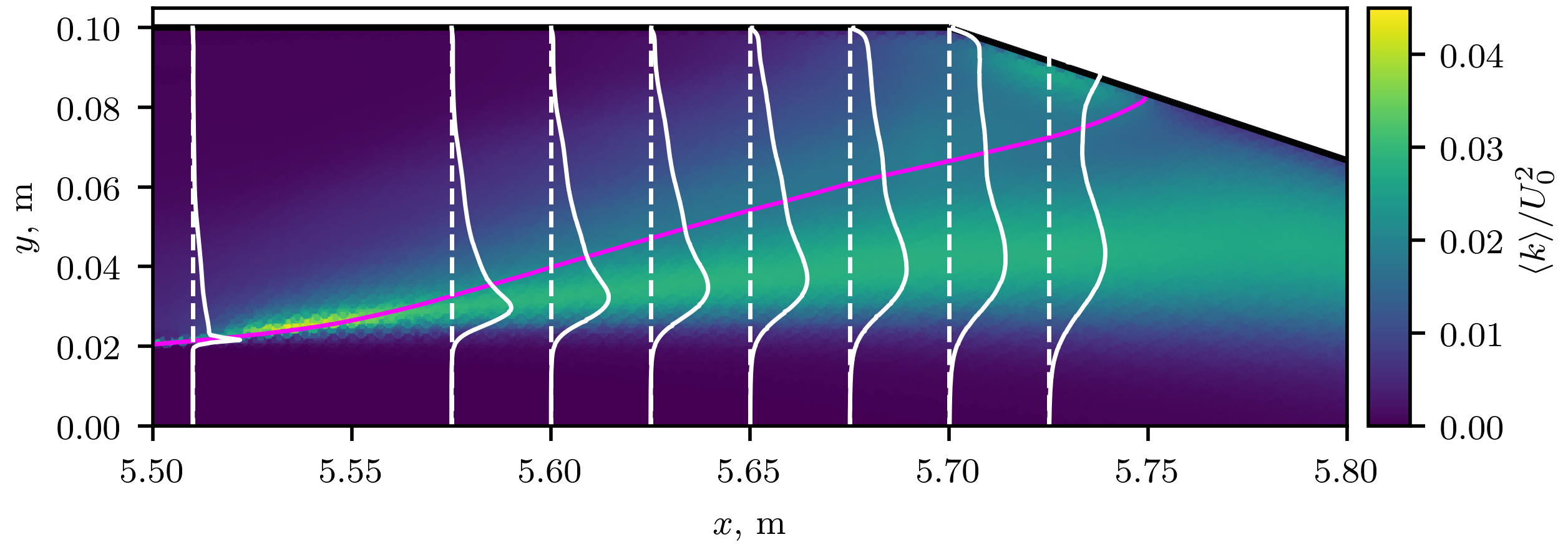}
	\caption
	{
		Distribution of $\mean{k}$ near the beach wall.
		White solid lines show profiles of $0.5\mean{k}/U^2_0$ at selected locations, marked by dashed white lines.
		The magenta line shows the  $\mean{\alpha_{0.5}}$ isoline.
	}
	\label{fig:k_2d}
\end{figure}

The discussion is continued with the analysis of the second-order moments of velocity.
In Figure~\ref{fig:k_2d}, the distribution of $\mean{k} = 0.5\mean{u'_iu'_i}$ is shown.
Interestingly, the peak of $\mean{k}$ occurs prior to the separation of the flow, in the region where the reverse flow plunges onto the steady wave.
Inspection of individual components of $\mean{u'_iu'_j}$ shows that this collision results in a large peak in $\mean{u'u'}$.
The detached shear layer is also turbulent, and  $\mean{k}$ clearly illustrates its growth in both directions along $y$.
The perturbation of the interface by the turbulent eddies in the shear layer facilitates mixing of air with water, contributing to the complicated state illustrated in Figure~\ref{fig:alpha_angle}.
Significant levels of turbulence are also observed in the air phase, but generally $\mean{k}$ decays above the mean position of the interface, except for a small region below the beach wall.


\begin{figure}[h!]
	\centering
	\includegraphics[]{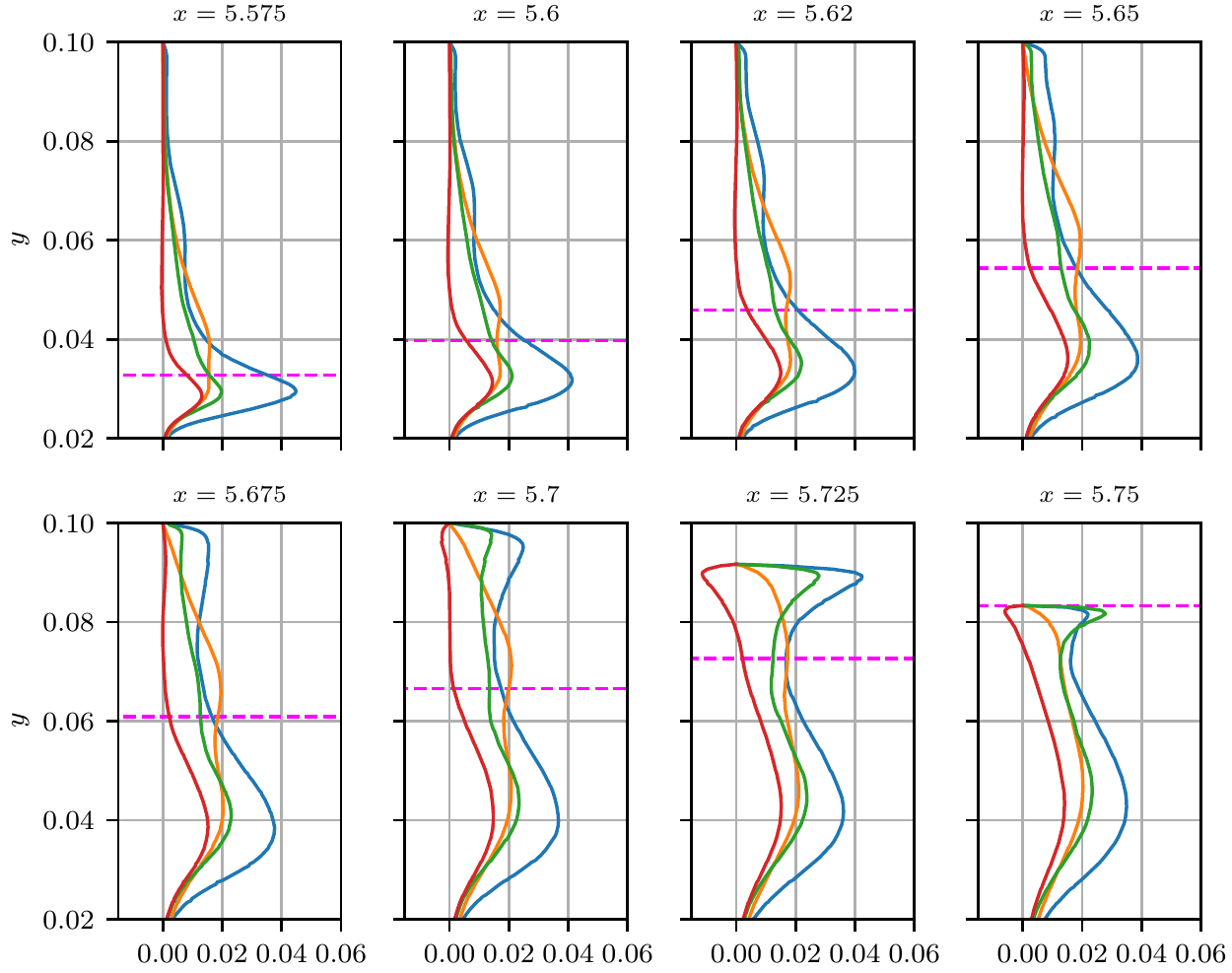}
	\caption
	{
		Profiles of the non-zero components of the $\mean{u'_iu'_j}$ tensor at selected values of $x$ close to the beach wall: $\mean{u'u'}/U_0^2$ --- blue line, $\mean{v'v'}/U_0^2$ --- orange line, $\mean{w'w'}/U_0^2$ --- green line, $\mean{u'v'}/U_0^2$ --- red line.
		The magenta dashed line shows the location of  the $\mean{\alpha_{0.5}}$ isoline.
	}
	\label{fig:rey}
\end{figure}

In Figure~\ref{fig:rey}, the non-zero components of $\mean{u'_iu'_j}$ are plotted at selected values of $x$ in the closure region.
Initially, the peaks in $\mean{u'_iu'_j}$ are found close to the interface, but further downstream their position shifts to about half-way between the interface and $y=0.02$, which is the approximate position of the separation point of the shear layer.
In terms of magnitude, the streamwise component $\mean{u'u'}$ is the largest, followed by $\mean{w'w'}$, $\mean{v'v'}$, and $\mean{u'v'}$.
Very similar observations have been made for the hydraulic jump flow~\citep{Mortazavi2016a}, further supporting its similarity with the flow in the air cavity's closure region.
Note that near the beach wall the sign of~$\mean{u'v'}$ changes to the opposite of that of the other components, which is typical of wall turbulence.

Time-signals of $u$ and $\alpha$ have been sampled at selected locations in order to compute energy spectra and spatial two-point correlations.
Both of these quantities can be used for assessing the quality of the conducted LES, which is particularly important here since rather diffusive numerical schemes have been used.
Eight $[x,y]$ locations for sampling the signals were chosen.
Four of them are close to mean location of the interface, and the other are inside the detached shear layer, see the red dots in Figure~\ref{fig:recirculation}.
At each $[x,y]$, the signal was sampled for a period of 10 s from each of the 200 cells in the spanwise direction.

\begin{figure}[h!]
	\centering
	\includegraphics[center]{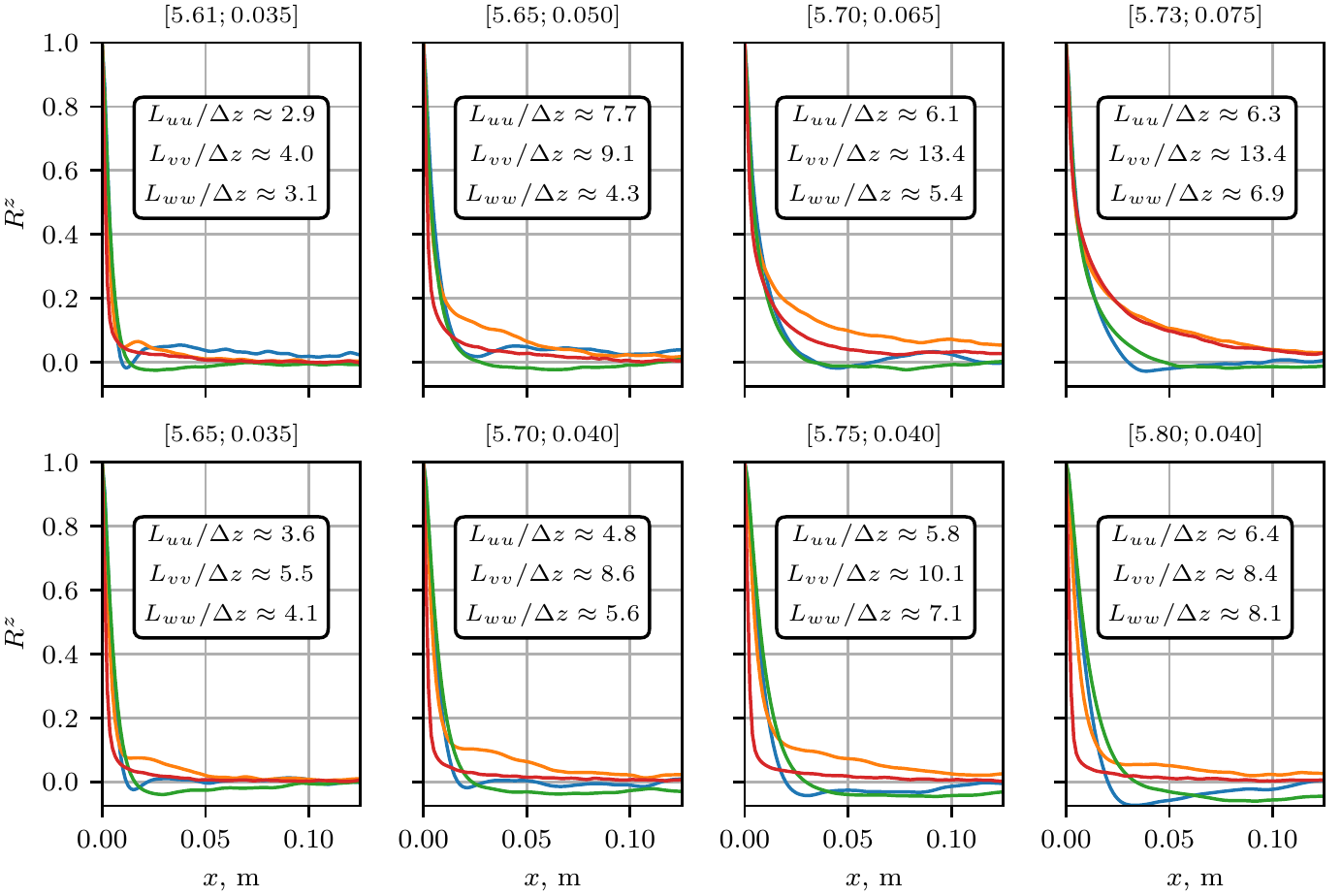}
	\caption
	{
		 Two-point autocorrelation functions in the spanwise direction.
		 $R_{uu}^z$ --- blue line, $R_{vv}^z$ --- orange line, $R_{ww}^z$ --- green line, $R_{\alpha \alpha}^z$ --- red line.
		 Plot titles indicate the $[x; y]$ coordinates for which the functions are computed.
	}
	\label{fig:2pcorr_z}
\end{figure}

Two-point spanwise autocorrelation functions of $\alpha$ and the velocity components are shown in Figure~\ref{fig:2pcorr_z}.
An important conclusion that can be drawn from the data is that the spanwise extent of the computational domain is sufficiently large in order for the cyclic boundary conditions to not introduce spurious periodicity effects.
Additionally, two-point correlations can be used to compute integral length scales $L^z_{u_i u_i}$, characterizing the size of the largest eddies.
The ratio of the integral length scales to the size of the cell in the corresponding direction is a reliable measure of the spatial resolution of the LES~\citep{Davidson2009}.
The ratios obtained for all three velocity components are presented in~Figure~\ref{fig:2pcorr_z} for each spatial location, with the mean value across all components and locations being $\approx 6.3$.
In~\citep{Davidson2009}, the recommended value for a coarse LES is 8, which is comparable.
It is also observed that in the detached shear layer, $L^z_{u_i u_i}$ increase with $x$, indicating that the turbulent structures become elongated along the spanwise direction.

\begin{figure}[h!]
	\centering
	\includegraphics[center]{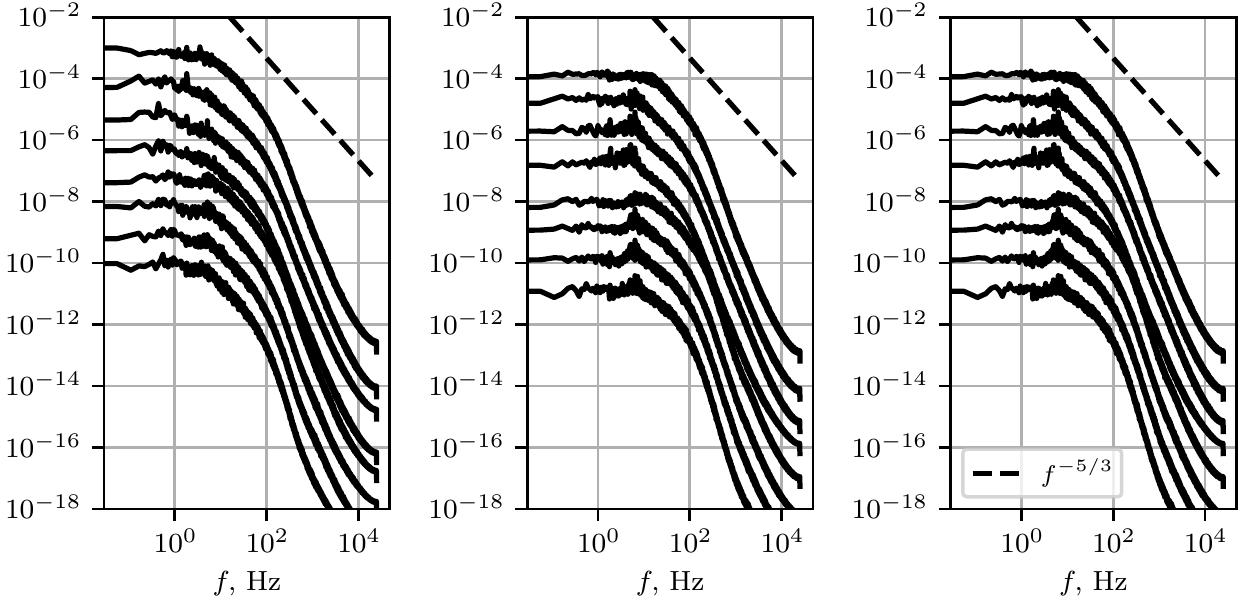}
	\caption
	{
		Temporal energy spectra of $u'$ --- left, $v'$ --- middle, $w'$ --- right.
		In each plot, different curves correspond to different $[x, y]$ locations.
		The data for each location is pre-multiplied with a different power of $10$ in order for the curves to be distinguishable.
		The order of the curves, from top to bottom, is the same as the consecutive order in Figure~\ref{fig:2pcorr_z}.
	}
	\label{fig:time_spectra}
\end{figure}

Figure~\ref{fig:time_spectra} shows the temporal energy spectra of the velocity components.
The energy levels are similar at all the considered $[x, y]$ locations, so in the plots the data from each location is premultiplied by a different factor of 10 to make the curves distinguishable.
The dashed line in the plot shows the $-5/3$ slope, which can be expected in the inertial subrange.
Parts of the spectra following this slope are distinctly seen in the spectra of $u'$ and $v'$, whereas for $w'$ the inertial range is less pronounced, in particular outside of the detached shear layer.
While the presence of the $-5/3$-range in the spectra is a poor indicator of LES quality~\citep{Davidson2009}, together with the data for two-point correlations discussed above it provides sufficient evidence that the conducted simulations can be considered a coarse LES.

\section{Conclusions} \label{sec:conclusions}

This paper reports the results of an LES of the flow in an internal air ship cavity.
Using this high-fidelity modelling approach, which was not previously applied to this type of flows, allowed to gain insight into the flow dynamics of the closure region of the cavity.
It is shown that the flow there is highly unsteady and turbulent, with a significant degree of mixing of air and water and periodically occurring shedding of air bubble swarms.
The analysis of the flow variables reveals that the stagnation of the free-stream flow on the beach wall gives rise to a pressure gradient, which pushes the air-water interface upwards.
As a result, flow separation occurs and a recirculation zone encompassing the interface is formed.
The  reverse flow driven by the recirculation facilitates  air entrainment and mixing, in particular, in the region where it plunges onto the steady wave.
Furthermore, the separated flow becomes turbulent, which enhances mixing and contributes to the highly irregular form of the interface.

Interestingly, several similarities between the flow in the closure region and that in a hydraulic jump have been observed.
This includes the separation from an inclining air-water interface, the behaviour of the Reynolds stresses as the separated flow develops, and the presence of a recirculation zone above it, which ultimately leads to the periodic air shedding.
This similarity can be used in order to develop computational methodologies using the much more extensively studied hydraulic jump flow and later apply them to internal air cavities.

At this point, it is unclear whether the established entrainment mechanism is dominant independent of the amplitude and angle of the gravity wave as it approaches the beach wall, as well as the design of the latter.
It is plausible that other mechanisms, such as those discussed in~\citep{Makiharju2010}, take precedence when the pressure gradient on the beach is not strong enough to significantly affect the location of the interface.
To establish whether this is the case, additional simulations are to be conducted.

Several other directions of future work can be identified.
One is to address the effect of the lubrication on the drag of the hull downstream of the cavity.
If a dramatic increase in computational cost is to be avoided, doing this would require introducing new types of modelling.
A potential approach is converting the leaked bubbles to Lagrangian particles.
Another topic of future investigation is developing lower-fidelity simulation approaches, which could reduce simulation times to levels that allow effectively exploring the design parameter space of the cavity.
It is anticipated that this would require introducing special modelling for air entrainment and large simulation campaigns to quantify the effect of associated model constants.
Finally, including the side-walls of the cavity in the simulation would allow to assess the effects of the three-dimensionality of the wave system on the stability of the cavity and the amount of air leakage.

\section{Acknowledgements}
This work was supported by grant number P38284-2 from the Swedish Energy Agency.
The simulations were performed on resources at Chalmers Centre for Computational Science and Engineering (C3SE).
The authors are grateful to Johan Roenby from Aalborg University for suggestions regarding using adaptive mesh refinement.
Gem Rotte from Delft University of Technology is gratefully acknowledged for many fruitful discussions on ship air cavity flows.

\bibliographystyle{plainnat}
\bibliography{library}

\appendix
\section{The influence of mesh resolution} \label{app:mesh}
In this appendix, a short study of the effects of the grid resolution on the simulation results is presented.
Recall that in order to construct the mesh, a coarse template-mesh was first generated, and then AMR was applied to refine it in the closure region and around the interface of the stationary wave upstream.
Furthermore, the level of refinement of the mesh was defined as the number of times a cell of the original template-mesh could be cut in half along all three spatial coordinates to produce eight cells in its place.
To produce the mesh used in the main simulation the refinement level was set to 3.
Three additional computations have been conducted on meshes corresponding to refinement level 0 (no refinement), 1, and 2.
The approximate number of cells in the meshes are, in ascending level of refinement, $7.57 \cdot 10^5$, $1.15 \cdot 10^6$, $4 \cdot 10^6$, and $25 \cdot 10^6$.

\begin{figure}[h!]
	\centering
	\includegraphics[width=0.45\linewidth]{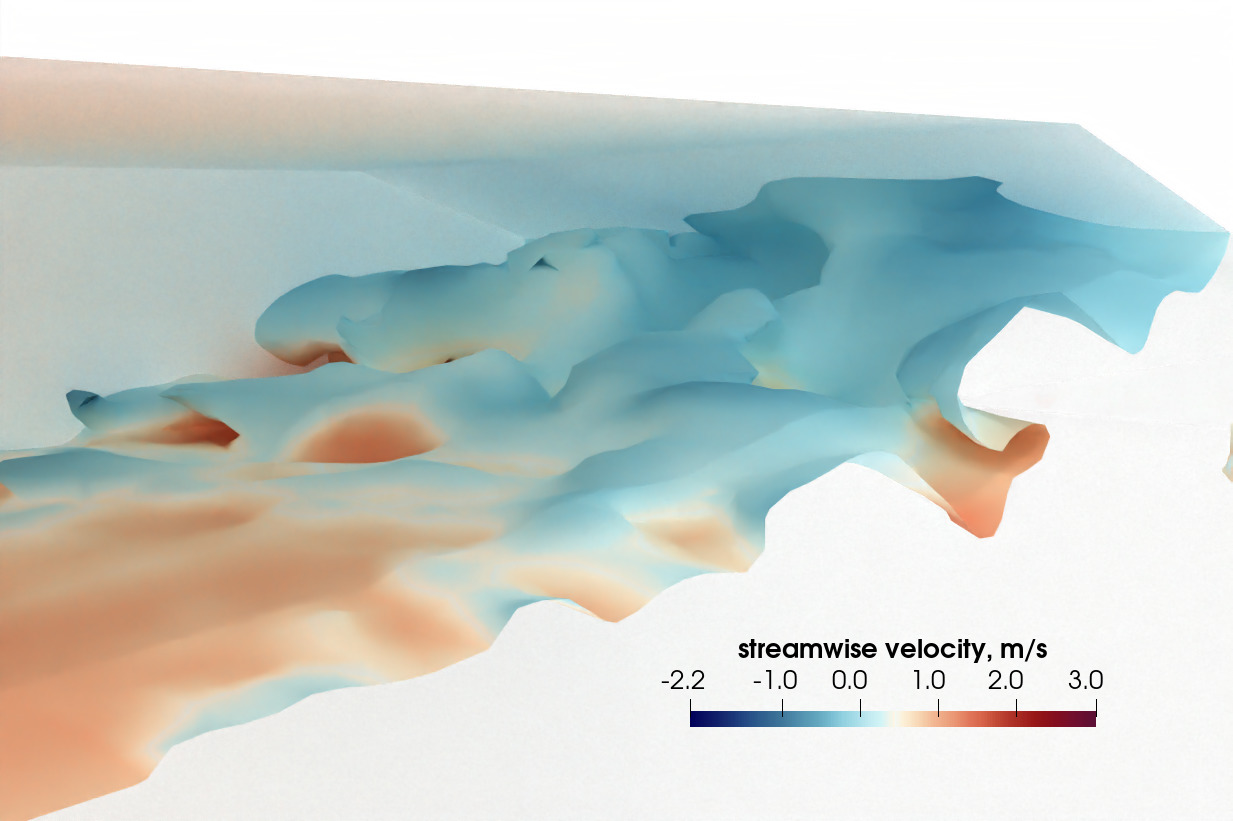}
	\includegraphics[width=0.45\linewidth]{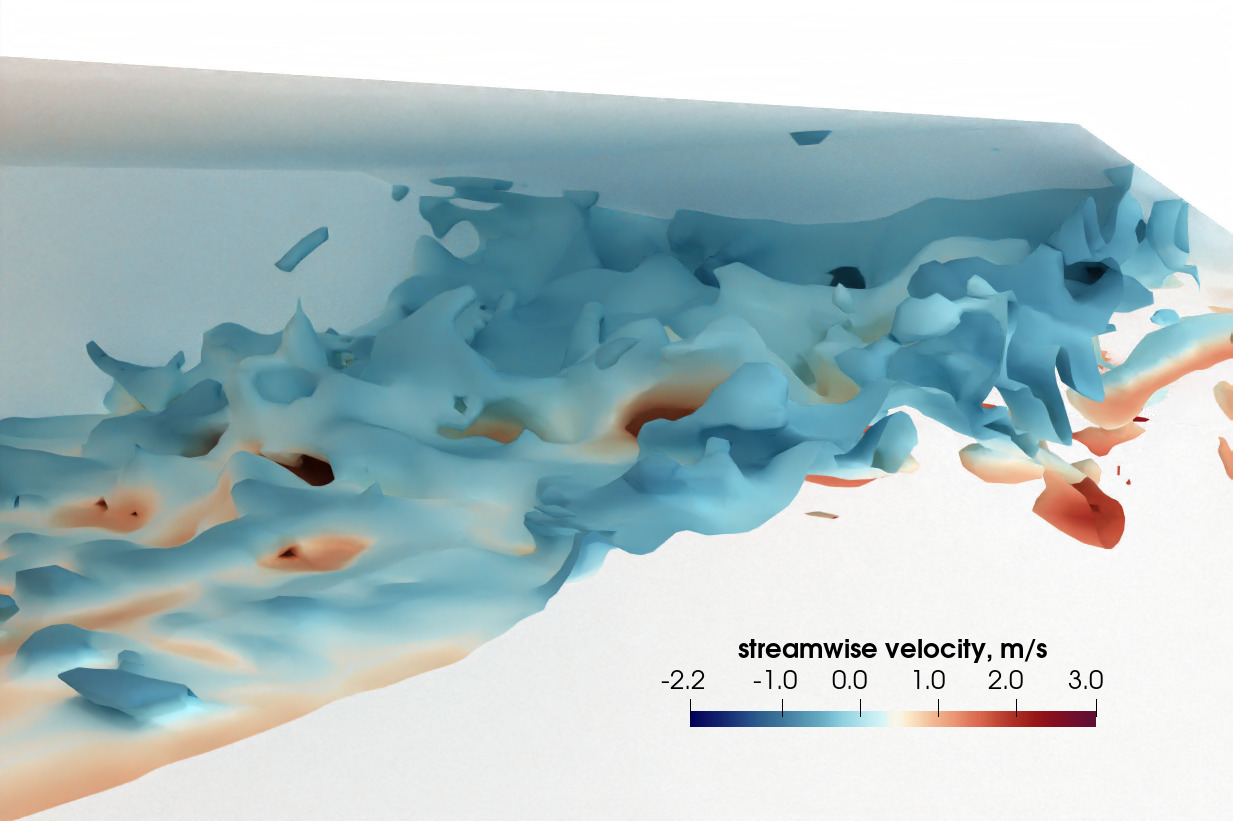}		\includegraphics[width=0.45\linewidth]{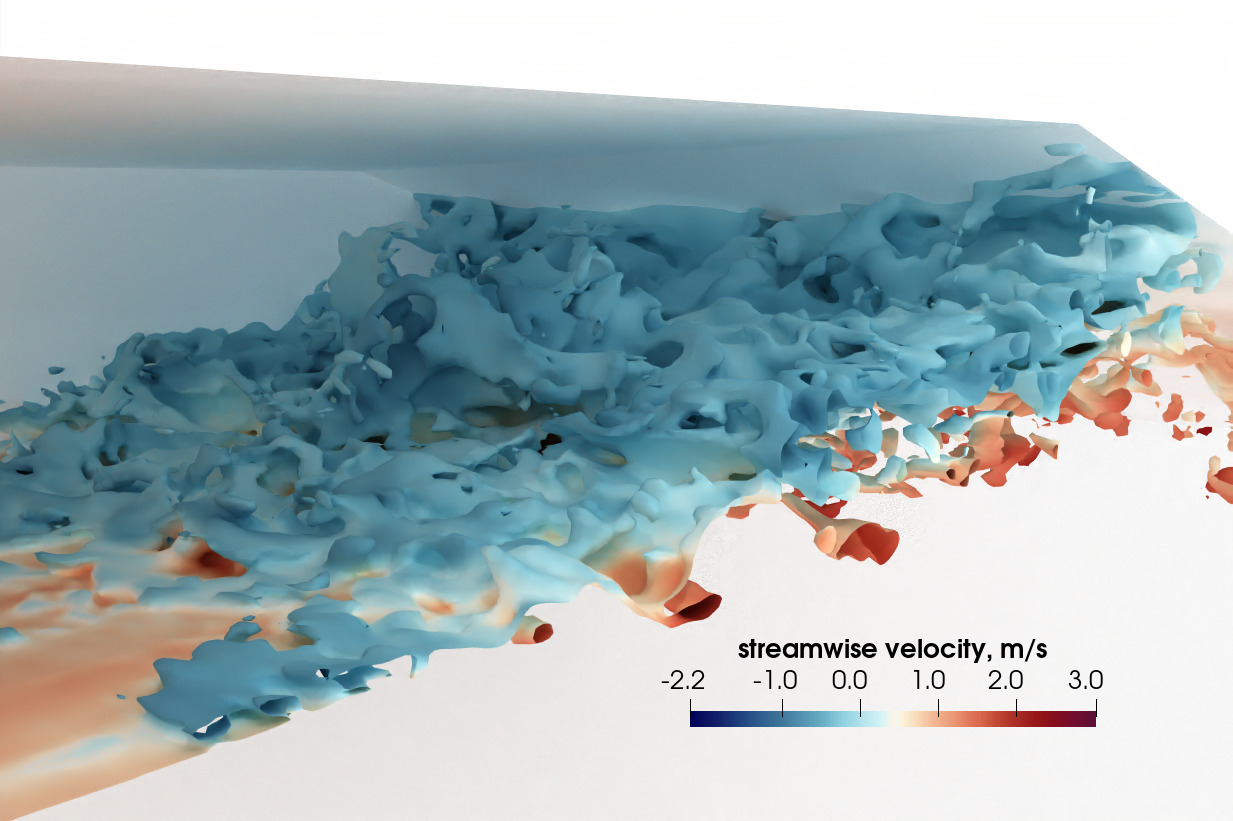}			\includegraphics[width=0.45\linewidth]{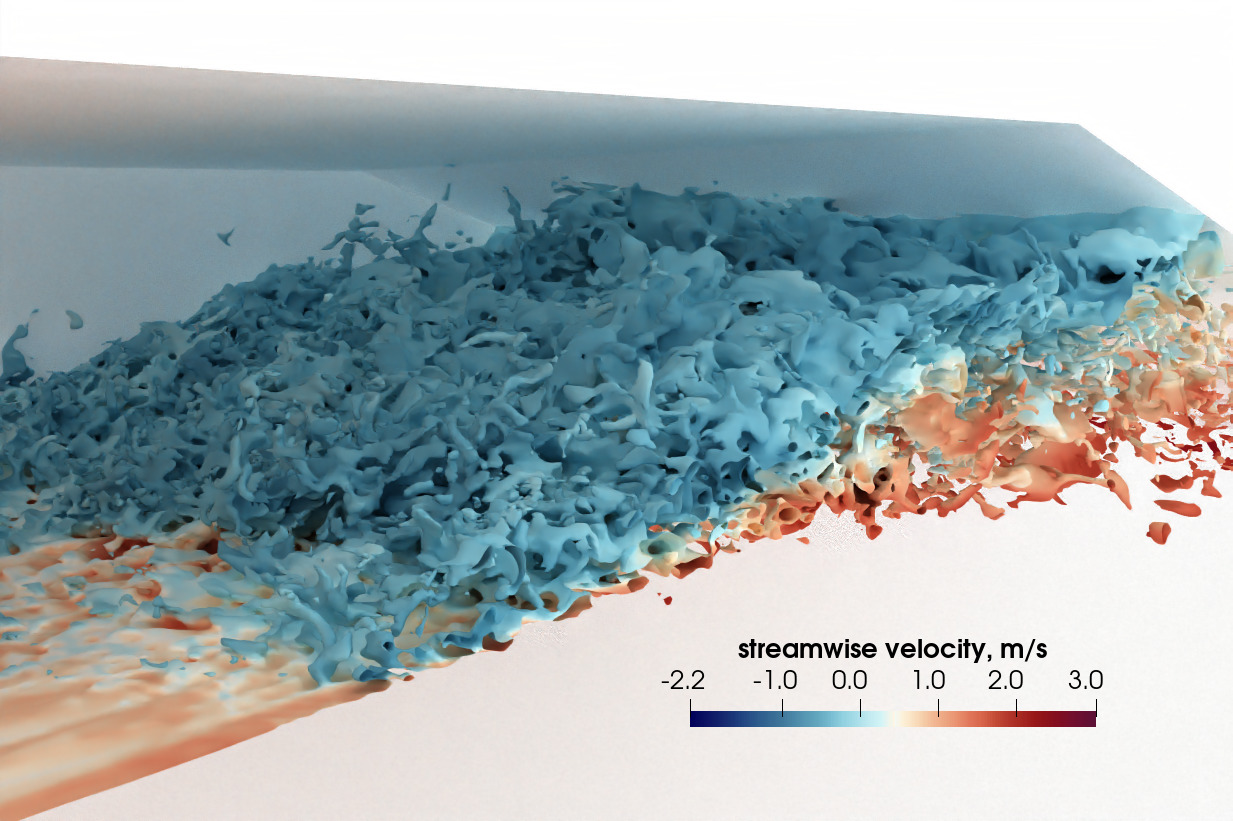}
	\caption
	{
		Snapshots of the $\alpha_{0.5}$ iso-surfaces obtained at different mesh refinement levels.
	}
	\label{fig:mesh_ref_alpha}
\end{figure}

A qualitative comparison between the levels of resolution achieved on each mesh can be made by considering the snapshots of the $\alpha_{0.5}$ isosurface, which are presented in Figure~\ref{fig:mesh_ref_alpha}.
The effects of grid resolution are highly evident, with structures of smaller sizes appearing in the solution with each consecutive refinement.

\begin{figure}[h!]
	\centering
	\includegraphics[]{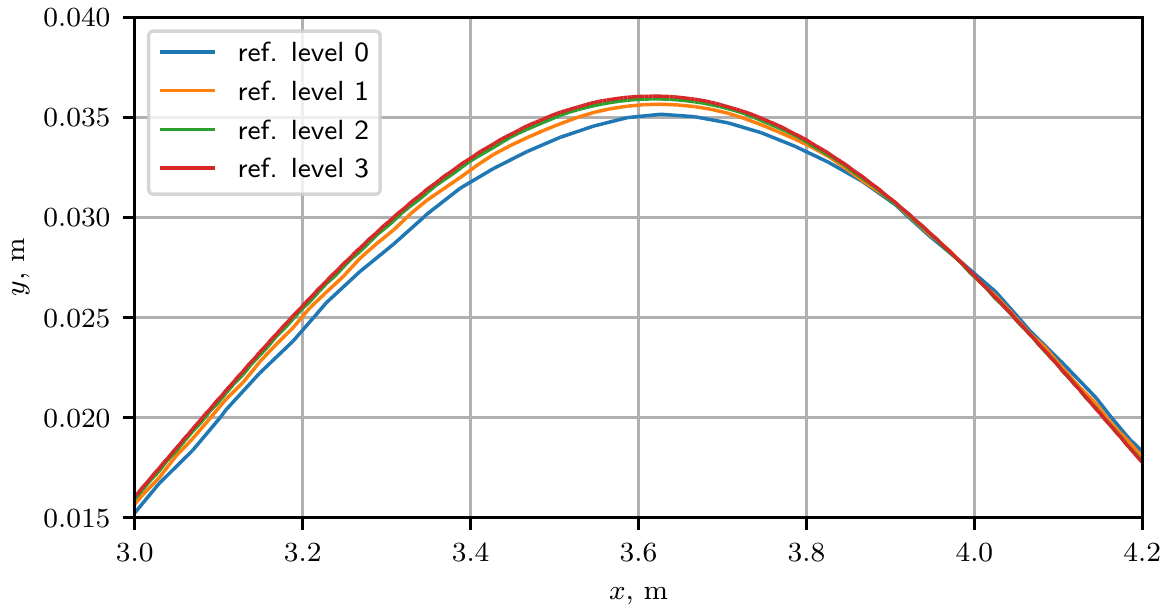}
	\caption
	{
		A close-up on the second crest of the stationary wave.
	}
	\label{fig:mesh_ref_wave}
\end{figure}

\begin{figure}[h!]
	\centering
	\includegraphics[]{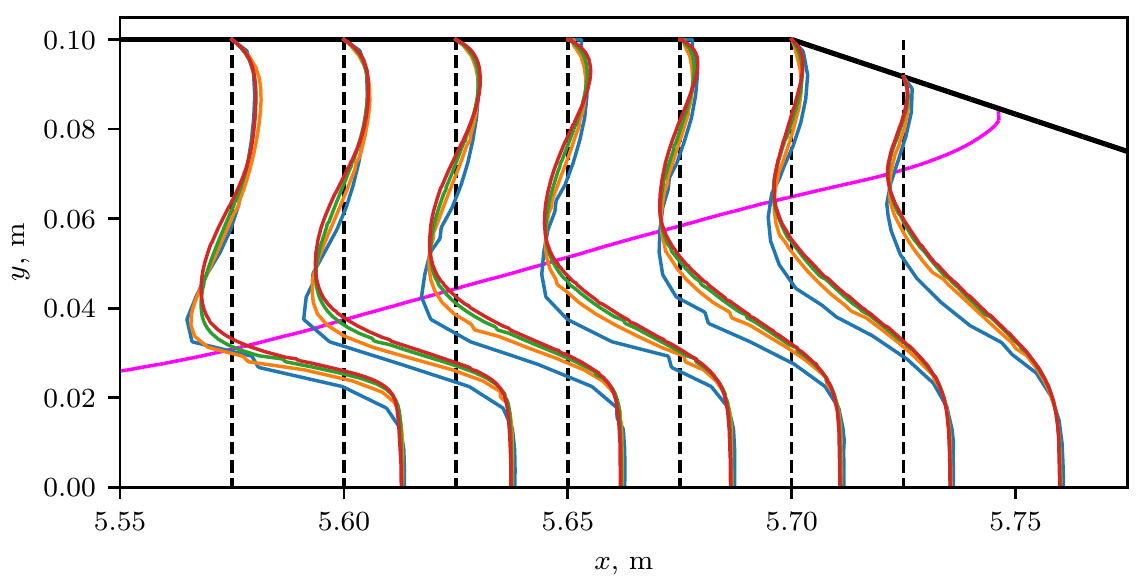}
	\caption
	{
		Streamwise velocity profiles at selected locations marked by vertical dashed lines. The magenta line shows the $\mean{\alpha_{0.5}}$ isoline, other line colours as in Figure~\ref{fig:mesh_ref_wave}.
	}
	\label{fig:mesh_ref_u}
\end{figure}

\begin{figure}[h!]
	\centering
	\includegraphics[]{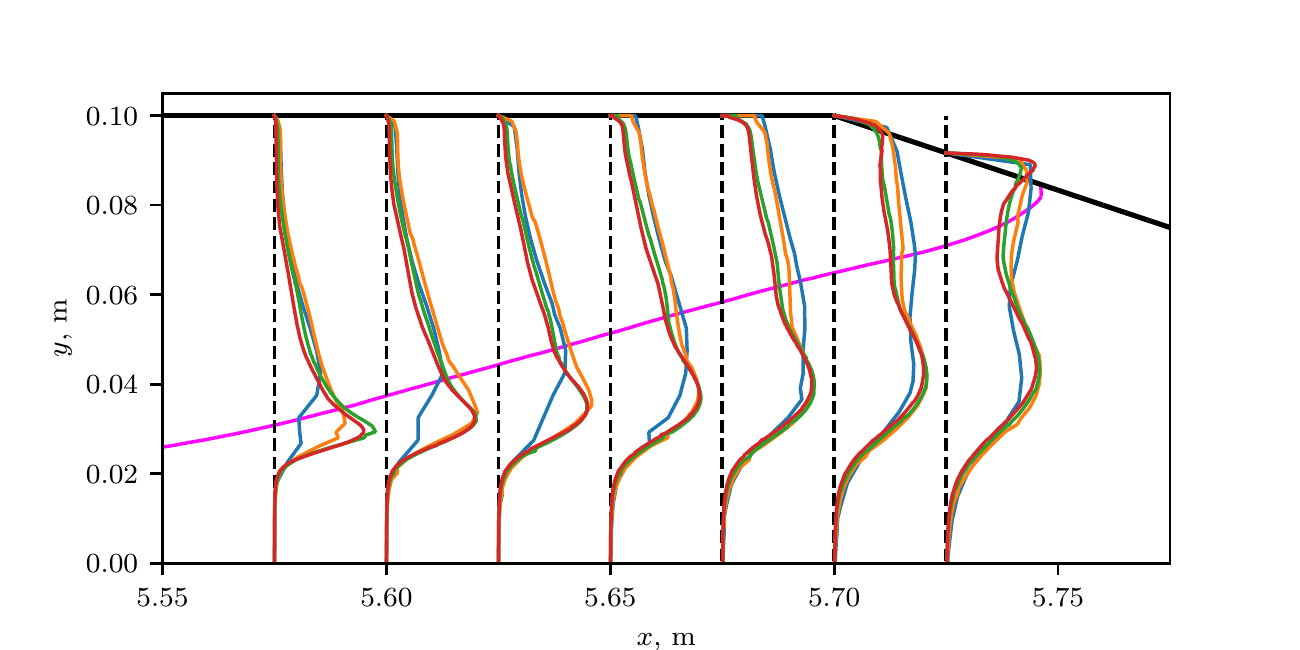}
	\caption
	{
	Profiles of $0.5\mean{k}/U^2_0$ at selected locations marked by vertical dashed lines. The magenta line shows the $\mean{\alpha_{0.5}}$ isoline, other line colours as in Figure~\ref{fig:mesh_ref_wave}.
	}
	\label{fig:mesh_ref_k}
\end{figure}

The differences in the obtained profiles of the stationary wave are so small that a close-up view is necessary to see them.
Figure~\ref{fig:mesh_ref_wave} shows a zoom-in on the second crest of the wave.
Figures~\ref{fig:mesh_ref_u} and~\ref{fig:mesh_ref_k} show, respectively, the profiles of $\mean u$ and $\mean k$.
For all of these quantities, the curves corresponding to refinement levels 2 and 3 are either indistinguishable or extremely close to each other.
\rev{This indicates that the mesh used in main simulation is sufficiently fine and no significant gain would be made from further mesh refinement}.
Furthermore, the results obtained on the mesh with refinement level 1 are also in good agreement with what is obtained in the more expensive simulations, even for $\mean k$.
The results on the coarsest mesh are also surprisingly good considering the fact that nearly no multiphase structures are resolved on this mesh as shown in Figure~\ref{fig:mesh_ref_alpha}.

In summary, considering the fact that the simulation on the densest mesh could only be classified as a coarse LES (see the end of Section~\ref{sec:results}), the accuracy of the results on the coarser meshes is very good.
Using the refinement level 2 mesh is possible with virtually no impediment to the predictive accuracy, and refinement level 1 is sufficient for quick evaluation of a new design or flow condition.
It is noted that the simulation time on the refinement level 1 mesh can be roughly approximated as one week on a modern workstation or a single node of a computational cluster, thus being easily affordable for industry.

\section{The influence of the length of the cavity} \label{app:cavity_length}

\begin{figure}[h!]
	\centering
	\includegraphics[]{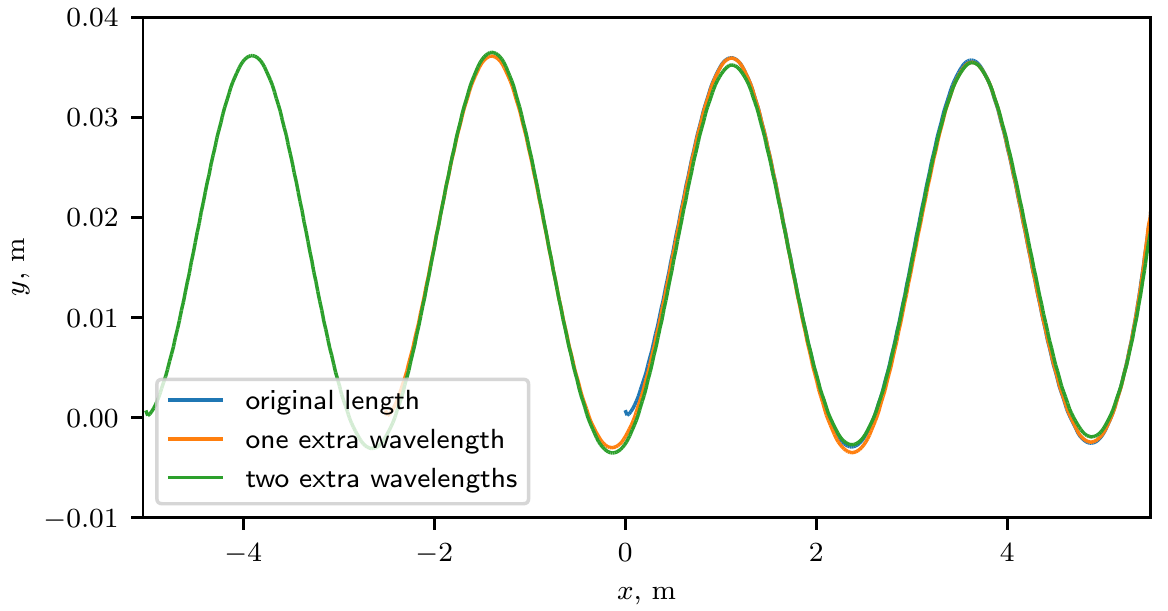}
	\caption
	{
		The stationary wave profiles obtained in simulations with three different cavity lengths.
	}
	\label{fig:cavity_length}
\end{figure}

Recall that a significant discrepancy has been observed between the value of $\eta_0$ obtained in the simulation and that predicted based on the height of the water column necessary to compensate the nominal pressure difference between the air in the cavity and the surrounding water.
If this discrepancy is explained by the influence of flow dynamics in the closure region, it is reasonable to expect that $\eta_0$ will be dependent on the length of the cavity.
To test this hypothesis, two additional simulations have been conducted, in which the length of the cavity has been increased by, respectively, one and two wave-lengths of the stationary wave.
This way the dynamics of the closure region remain unaltered.
The simulations were performed on a mesh with the AMR refinement level set to~1.
The obtained profiles of the waves are presented in Figure~\ref{fig:cavity_length}.
It is clear that the influence of the cavity length on $\eta_0$ is negligible.

\section{The influence of surface tension} \label{app:tension}

\begin{figure}[h!]
	\centering
	\includegraphics[]{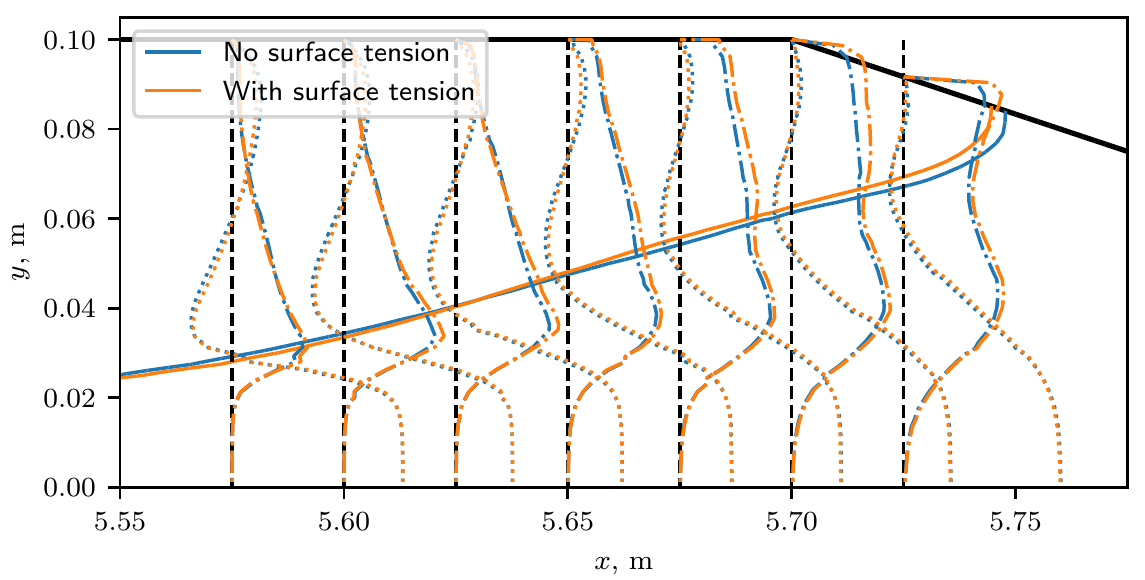}
	\caption
	{
		Comparison between simulations with and without surface tension modelling.
		\textit{Solid lines}: $\mean{\alpha_{0.5}}$ isolines. \textit{Dotted lines:} profiles of $\mean{u}/(25U_0)$. \textit{Dashed-dotted lines: $0.5\mean{k}/U^2_0$}.
	}
	\label{fig:surface_tension}
\end{figure}

As discussed in Section~\ref{sec:cfd}, the main simulation has been conducted without considering the effects of surface tension.
To account for it, the Continuum Force Model can be employed, in which the following term is introduced into the right-hand-side of the momentum equation~\eqref{eq:lesmom}:

\begin{equation}
\label{eq:surface_tension}
f^s_i = \gamma \kappa \pdiff{\alpha}{x_i},
\end{equation}
where $\gamma = 0.07$ N/m is the surface tension coefficient and $\kappa = \partial n^f_i/\partial x_i$ is the curvature of the interface between the two phases, where $n^f_i$ is the interface unit-normal, see~\eqref{eq:normal}.
Recall that this model introduces parasitic currents into the solution, which destabilizes the simulation, and is the reason why it was not used.
However, it was possible to conduct a simulation with surface tension modelling included on a coarser mesh obtained by setting the AMR refinement level to~1.
A comparison between the results with and without surface tension modelling is presented in Figure~\ref{fig:surface_tension}.
It is evident that the effect of surface tension is small enough to neglect it.
Nevertheless, it can be argued that surface tension will become more important at higher resolutions since it will prevent the break-up of small bubbles.
But on the other hand, the conducted study on the effects of mesh resolution (see Appendix~\ref{app:mesh}) seems to indicate that small bubbles do not have a strong influence on the flow.

\printcredits

\end{document}